\documentclass[journal]{IEEEtran}

\usepackage[dvipsnames]{xcolor}
\usepackage{amsmath}
\usepackage{mathrsfs}
\usepackage{amssymb}
\usepackage[mathcal]{euscript}
\usepackage{graphicx}
\usepackage{psfrag}
\usepackage{subfigure}
\usepackage{url}
\usepackage{stfloats}
\usepackage{amsmath}
\usepackage{pifont}  
\usepackage{graphicx} 
\usepackage{float}
\usepackage{array}
\usepackage{algorithm}
\usepackage{amsthm}
\usepackage{booktabs} 
\usepackage{algorithm} 
\usepackage{algorithmic} 
\usepackage{color}
\usepackage{cases}
\usepackage{bm}
\usepackage{amsmath,amsfonts,amssymb,amsbsy,paralist,ifthen}

\newcommand{\cmark}{\ding{51}} 
\newcommand{\xmark}{\ding{55}} 

\makeatletter
\def\UrlAlphabet{%
      \do\a\do\b\do\c\do\d\do\e\do\f\do\g\do\h\do\i\do\j%
      \do\k\do\l\do\m\do\n\do\o\do\p\do\q\do\r\do\s\do\t%
      \do\u\do\v\do\w\do\x\do\y\do\z\do\A\do\B\do\C\do\D%
      \do\E\do\F\do\G\do\H\do\I\do\J\do\K\do\L\do\M\do\N%
      \do\O\do\P\do\Q\do\R\do\S\do\T\do\U\do\V\do\W\do\X%
      \do\Y\do\Z}
\def\UrlDigits{\do\1\do\2\do\3\do\4\do\5\do\6\do\7\do\8\do\9\do\0}
\g@addto@macro{\UrlBreaks}{\UrlOrds}
\g@addto@macro{\UrlBreaks}{\UrlAlphabet}
\g@addto@macro{\UrlBreaks}{\UrlDigits}
\makeatother

\usepackage{colortbl}
\usepackage{arydshln}
\usepackage{multirow}
\usepackage{tabularray}
\usepackage{multicol}
\usepackage{tabularx}

\usepackage{color}

\usepackage{pifont}

\newcolumntype{M}[1]{>{\centering\arraybackslash}m{#1}} 
\newcolumntype{J}[1]{>{\raggedright\arraybackslash}m{#1}} 

\begin{document}
%

\title{Toward Edge General Intelligence with Agentic AI and Agentification: Concepts, Technologies, and Future Directions}

\author{Ruichen Zhang, Guangyuan Liu, Yinqiu Liu, Changyuan Zhao, Jiacheng Wang, Yunting Xu,   \\
Dusit Niyato,~\IEEEmembership{Fellow,~IEEE}, Jiawen Kang,  Yonghui Li,~\IEEEmembership{Fellow,~IEEE},  Shiwen Mao,~\IEEEmembership{Fellow,~IEEE}, \\ Sumei Sun,~\IEEEmembership{Fellow,~IEEE},  Xuemin Shen,~\IEEEmembership{Fellow,~IEEE}, and Dong In Kim,~\IEEEmembership{Life Fellow,~IEEE}

\thanks{R. Zhang, G. Liu, Y. Liu, C. Zhao, J. Wang, Y. Xu, and D. Niyato are with the College of Computing and Data Science, Nanyang Technological University, Singapore (e-mail: ruichen.zhang@ntu.edu.sg, liug0022@e.ntu.edu.sg,  yinqiu001@e.ntu.edu.sg,  zhao0441@e.ntu.edu.sg, jiacheng.wang@ntu.edu.sg, yunting.xu@ntu.edu.sg, dniyato@ntu.edu.sg).}

\thanks{J. Kang is with the School of Automation, Guangdong University of Technology, Guangzhou 510006, China (e-mail: kavinkang@gdut.edu.cn).}

\thanks{Y. Li and is with the School of Electrical and Information Engineering, University of Sydney, Sydney, NSW 2006, Australia (e-mail: yonghui.li@sydney.edu.au).}

\thanks{S. Mao is with the Department of Electrical and Computer Engineering,
Auburn University, Auburn, USA (e-mail: smao@ieee.org).}

\thanks{S. Sun is with the Institute for Infocomm Research, Agency for Science, Technology and Research, Singapore (e-mail: sunsm@i2r.a-star.edu.sg).} 

\thanks{X. Shen is with the Department of Electrical
and Computer Engineering, University of Waterloo, Waterloo, ON N2L 3G1, Canada (e-mail: sshen@uwaterloo.ca).}

\thanks{D. I. Kim is with the Department of
Electrical and Computer Engineering, Sungkyunkwan University, Suwon 16419, South Korea (email:dongin@skku.edu).}

}

\maketitle

\begin{abstract}
The rapid expansion of sixth-generation (6G) wireless networks and the Internet of Things (IoT) has catalyzed the evolution from centralized cloud intelligence towards decentralized edge general intelligence. However, traditional edge intelligence methods, characterized by static models and limited cognitive autonomy, fail to address the dynamic, heterogeneous, and resource-constrained scenarios inherent to emerging edge networks. Agentic artificial intelligence (Agentic AI) emerges as a transformative solution, enabling edge systems to autonomously perceive multimodal environments, reason contextually, and adapt proactively through continuous perception–reasoning–action loops. In this context, the agentification of edge intelligence serves as a key paradigm shift, where distributed entities evolve into autonomous agents capable of collaboration and continual adaptation. This paper presents a comprehensive survey dedicated to Agentic AI and agentification frameworks tailored explicitly for edge general intelligence. First, we systematically introduce foundational concepts and clarify distinctions from traditional edge intelligence paradigms. Second, we analyze important enabling technologies, including compact model compression, energy-aware computing strategies, robust connectivity frameworks, and advanced knowledge representation and reasoning mechanisms. Third, we provide representative case studies demonstrating Agentic AI’s capabilities in low-altitude economy networks, intent-driven networking, vehicular networks, and human-centric service provisioning, supported by numerical evaluations. Furthermore, we identify current research challenges, review emerging open-source platforms, and highlight promising future research directions to guide robust, scalable, and trustworthy Agentic AI deployments for next-generation edge environments.
\end{abstract}



\section{Introduction}

\subsection{Background}

The rollout of sixth-generation (6G) wireless networks is ushering in a transformative era driven by the rapid expansion of edge-connected devices~\cite{peltonen20206gwhitepaperedge,10955732,wu2023split,9023459}. According to IoT Analytics, the number of globally connected IoT devices will reach approximately {27.1 billion} by 2025, increasing significantly from 16.6 billion in 2023\footnote{\url{https://iot-analytics.com/number-connected-iot-devices/}}. Concurrently, Gartner forecasts that roughly {75\% of enterprise-generated data} will be processed at the network edge by 2025\footnote{\url{www.gartner.com/smarterwithgartner/what-edge-computing-means-for-infrastructure-and-operations-leaders}}. This explosive growth in edge connectivity has catalyzed a fundamental paradigm shift, moving intelligence from centralized cloud infrastructures toward decentralized edge intelligence~\cite{10681129,hu2023adaptive}. Edge intelligence has become integral to latency-sensitive and mission-critical applications, such as autonomous driving, industrial automation, drone swarms, and real-time healthcare monitoring, where centralized cloud solutions fail to meet stringent latency, reliability, and privacy requirements~\cite{10879580,10213996,11021376}.

Despite its substantial benefits, traditional edge intelligence methods are predominantly based on static, task-specific models, designed primarily for single or narrowly defined tasks such as object detection or simple predictive analytics~\cite{10304612,xu2021edge,9961954,7438736}. For instance, classical ei-based drone swarms often employ fixed, predefined trajectory plans, unable to effectively adapt to sudden mission changes or environmental disturbances~\cite{Cao2020IoT,10037234,9628162}. This inflexibility significantly limits operational effectiveness and safety in dynamic, multi-modal edge environments. To overcome these limitations, the concept of {edge general intelligence} has emerged~\cite{10909637,9786719,10089482}. Edge general intelligence integrates broader cognitive capabilities, including multi-task generalization, continual learning, and contextual understanding, directly into edge devices, empowering them to autonomously navigate complex and evolving operational scenarios~\cite{Shen2024EI,10945973,11059885}. For example, edge general intelligence enables smart grid systems to dynamically adjust energy distribution under fluctuating loads, supports autonomous drone swarms in real-time mission adaptation, and facilitates seamless human–robot collaboration in dynamically changing factory environments~\cite{zhang2025model,shen2021holistic}.

Nevertheless, implementing edge general intelligence presents several formidable challenges. Edge devices often operate under stringent constraints on computational power, memory capacity, and energy availability~\cite{Zhu2022TGCN,8123913,8847416}. Furthermore, effective real-time scalability, robust multi-modal data processing, and long-horizon reasoning capabilities remain significant hurdles for current edge intelligence frameworks~\cite{liu2025advances,8612450},li2021slicing. These inherent limitations necessitate a paradigm shift toward more robust, adaptive, and autonomous forms of intelligence, capable of fully leveraging edge-device potentials~\cite{He2024TMC,feng2025multiagentembodiedaiadvances}.

In response to these critical challenges, Agentic artificial intelligence (Agentic AI) emerges as a transformative paradigm, fundamentally redefining the capabilities of edge intelligence~\cite{10506539,10811956}. \textit{Agentic AI refers to intelligent systems characterized by continuous perception–reasoning–action loops, enabling autonomous context interpretation, explicit reasoning, and goal-driven decision-making~\cite{tong2025wirelessagent}.} This process of agentification marks a shift from passive entities toward autonomous, adaptive agents, empowering edge systems with greater cognitive autonomy and coordination \cite{pico2018agentification}. Unlike conventional edge intelligence that operates with static inference pipelines, Agentic AI leverages advanced generative frameworks, particularly large language models (LLMs), to dynamically integrate multi-modal data, perform deliberative planning, and autonomously orchestrate decentralized tasks~\cite{10815045,10835069,luo2025weighted}. Such capabilities, oftentimes referred to as agentification, allow edge devices to not only respond reactively, but also anticipate proactively, reason about, and adapt to complex environments~\cite{9462495,luo2025edgegeneralintelligencemultiplelarge}.

\begin{table*}[!t]
\caption{SUMMARY OF Related WORKS} 
\label{tab:agentic-edge-wireless}
\centering
\footnotesize
\setlength{\tabcolsep}{4pt}
\renewcommand{\arraystretch}{1.15}
\begin{tabularx}{\textwidth}{l|X|l|c|c|c}
\hline
\textbf{Ref.} & \textbf{Overview} & \textbf{Type} & \textbf{Agentic AI} & \textbf{Edge Intelligence} & \textbf{Wireless Networks} \\
\hline
\cite{sapkota2025uavs} &
A multidomain survey of agentic UAVs integrating perception, memory, decision-making, and collaborative planning, mapping application domains and roadmaps for autonomous aerial ecosystems &
Survey & \color{green}\cmark & \color{red}\xmark & \color{red}\xmark \\
\hline
\cite{11085101} &
An article proposing an edge large ai model-empowered cognitive multimodal semantic communication agent that performs intent understanding and planning-based policy generation&
Journal & \color{red}\xmark & \color{green}\cmark & \color{green}\cmark \\ 
\hline
\cite{dev2025advanced} &
An overview of advanced 6G architectures integrating agentic AI, constrained-AI operations, serverless orchestration, and optical low-latency fabrics to cut operational expenditure and enable new services &
Magazine & \color{green}\cmark & \color{red}\xmark & \color{green}\cmark \\
\hline
\cite{jiang2025large} &
A tutorial tracing the evolution from large AI models to agentic AI for intelligent communications, detailing core components (planner, tools, memory and knowledge base), multi-agent systems, and representative 6G applications &
Tutorial & \color{green}\cmark & \color{red}\xmark & \color{green}\cmark \\
\hline
\cite{xiao2025towards} &
An article proposing a generative foundation model-as-agent framework that supports interaction, collaborative learning, and knowledge transfer among agents for 6G networking, illustrated with digital-twin and metaverse scenarios&
Journal & \color{green}\cmark & \color{red}\xmark & \color{green}\cmark \\
\hline
\cite{tong2025wirelessagent} &
An article introducing LLM-based agents with perception, memory, planning, and action for wireless tasks such as network slicing, achieving near-optimal throughput across diverse scenarios &
Journal & \color{green}\cmark & \color{red}\xmark & \color{green}\cmark \\
\hline
\cite{salama2025edge} &
An edge agentic AI framework integrated into the O\mbox{-}Radio Access Network Intelligent Controller that combines persona-based multi-tool agents, predictive anomaly detection, and safety-aligned rewards for autonomous network optimization &
Journal & \color{green}\cmark & \textit{Partially} & \color{green}\cmark \\
\hline
\cite{10811956} &
A magazine article proposing Agent-as-a-Service, an AI-native edge framework in which agents plan, orchestrate, and manage 6G edge tasks via deviceless computing and webassembly &
Magazine & \textit{Partially} & \color{green}\cmark & \color{green}\cmark \\
\hline
\cite{lu2025agentic} &
A comprehensive survey introducing agentic to organize graph neural networks for scenario and task-aware wireless design, reviewing network applications (reconfigurable intelligent surface and cell-free) toward edge general intelligence &
Survey & \textit{Partially} & \color{green}\cmark & \color{green}\cmark \\
\hline
\textbf{\textit{Ours}} &
A comprehensive survey on Agentic AI frameworks for edge intelligence, introducing enabling technologies, representative case studies, and future directions toward scalable and trustworthy deployments in next-generation wireless edge networks&
Survey+Tutorial & \color{green}\cmark  & \color{green}\cmark & \color{green}\cmark \\
\hline
\end{tabularx}
\end{table*}

Recent innovative frameworks exemplify Agentic AI's profound potential. For instance, AutoGPT autonomously decomposes complex objectives into executable subtasks, dynamically orchestrating decentralized toolchains to achieve sophisticated goals such as network resource optimization and real-time robotic missions~\cite{Significant_Gravitas_AutoGPT}. Similarly, Voyager integrates long-horizon planning, contextual memory, and iterative self-improvement mechanisms, significantly enhancing UAV swarm coordination and environmental exploration capabilities~\cite{wang2023voyager}. Furthermore, Agentic AI has shown superior performance in vehicular networks, enabling autonomous vehicles to collaboratively adapt to dynamic traffic scenarios~\cite{10213996,10833743}, and in smart manufacturing environments, facilitating dynamic scheduling, predictive maintenance, and effective human–robot collaboration~\cite{8658105,11071266}. These practical examples vividly demonstrate how Agentic AI significantly surpasses traditional edge intelligence paradigms in adaptability, generalization, and operational intelligence, thereby becoming indispensable for next-generation edge deployments~\cite{gill2024edgeaitaxonomysystematic,zhang2025agenticaigenerativeinformation}.

\subsection{Motivation and Contributions}





{Despite the clear potential and early successes of Agentic AI and its agentification process, comprehensive exploration and systematic deployment methodologies tailored explicitly for multi-modal Agentic AI in 6G-enabled edge environments remain limited~\cite{jiang2025large,10945362,wang2023applicationsexplainableai6g}. Traditional AI architectures, such as static LLMs~\cite{brown2020language}, mixture-of-experts (MoEs)~\cite{shazeer2017outrageously}, foundation models~\cite{bommasani2021opportunities}, and embodied AI frameworks~\cite{ahn2024autortembodiedfoundationmodels} typically rely on one-way inference without fully considering the perception–reasoning–action loop~\cite{10908560,10720847,9906430}. Consequently, they lack sufficient adaptability, multi-modal integration, and cognitive autonomy necessary for robust operation in complex, evolving edge environments~\cite{10756738,10474509}. This limitation motivates a deeper, systematic study into the key design principles required to deploy practical and scalable Agentic AI systems~\cite{tallam2025autonomousagentsintegratedsystems}.}

{Table~\ref{tab:agentic-edge-wireless} summarizes the existing works, which address certain aspects of multi-modal Agentic AI; however, a unified methodology tailored for edge-oriented scenarios has yet to be established. Although many studies have explored Agentic AI across UAV autonomy, 6G architectures, and network control, most of them focused on isolated components or use cases rather than a coherent deployment framework for edge environments. For example, Sapkota \textit{et al.}~\cite{sapkota2025uavs} offered a multidomain survey of \emph{agentic UAVs} with rich autonomy, yet their scope did not systematize edge intelligence or generic wireless architectures. At the application layer, Sun \textit{et al.}~\cite{11085101} and Tong \textit{et al.}~\cite{tong2025wirelessagent} designed task-specific LLM-based agents for wireless tasks; however, these were not surveys and did not extract deployment methodologies for resource-constrained edges. At the architectural layer, Dev \textit{et al.}~\cite{dev2025advanced}, Xiao \textit{et al.}~\cite{xiao2025towards}, and Li \textit{et al.} (i.e., Agent-as-a-Service)~\cite{10811956} discussed 6G frameworks that incorporated agentic elements, but they did not cover a tutorial-style design flow for multi-modal agents under tight edge constraints. O\mbox{-}RAN centric work by Salama \textit{et al.}~\cite{salama2025edge} bridged toward practical RAN control but only partially addressed edge intelligence, while Lu \textit{et al.}~\cite{lu2025agentic} surveyed “agentic” graph neural networks from a graph-learning perspective rather than a general agent stack for edge general intelligence. Complementary tutorials such as Jiang \textit{et al.}~\cite{jiang2025large} traced the evolution toward agentic AI for communications; however, an integrative perspective on edge general intelligence that links agent capabilities to the networking stack, deployment sites (i.e., device, edge and cloud), and open toolchains remains underdeveloped.}

{In this survey, we aim to systematically explore critical design pillars and provide a comprehensive understanding of Agentic AI in the context of edge general intelligence. Differing from prior works summarized in Table~\ref{tab:agentic-edge-wireless}, which primarily examined isolated components, single-application agents, or high-level 6G frameworks, we articulate an \emph{edge-oriented and multi-modal} methodology that links agent capabilities to the wireless networking stack, concrete deployment sites (device, edge, cloud), and reproducible system stacks with metrics and benchmarks. Specifically, we identify four foundational design principles that underpin effective Agentic AI deployment at the edge:}
\begin{itemize}
    \item \textbf{Compactness:} Developing lightweight Agentic AI models and their agentification processes that are resource-efficient enough to run on edge devices with strict energy and hardware constraints, while retaining sufficient cognitive expressiveness and autonomy~\cite{zhang2025agenticaigenerativeinformation,xu2025decentralizationgenerativeaimixture}. Typical pathways include small language models with Low-Rank Adaptation of LLMs (LoRA), distillation, and quantization, evaluated by parameter count, memory footprint, multiply–accumulate operations, and energy per inference.
    
    \item \textbf{Efficiency:} Ensuring real-time responsiveness through computationally efficient inference and communication-aware collaborative protocols that meet stringent latency and reliability requirements of edge environments~\cite{10879580,9606720}. Key mechanisms include early-exit inference, approximate decoding, task offloading, and bandwidth-conscious coordination; representative metrics include end-to-end latency, reliability, throughput, and cost under service-level targets.
    
    \item \textbf{Knowledge and Reasoning:} Incorporating explicit, interpretable, and context-sensitive reasoning capabilities, enabling agents to handle complex, long-horizon decision-making scenarios with confidence and transparency~\cite{wang2023augmenting,yao2023react}. Practical enablers include structured memory, retrieval-augmented generation, and tool use, assessed by task success, reasoning faithfulness, retrieval consistency, and explanation quality.
    
    \item \textbf{Migration:} Facilitating seamless transfer and reuse of knowledge, skills, and tasks across diverse and dynamically changing network conditions, enhancing generalization, robustness, and reducing retraining overhead~\cite{zhang2024optimizinggenerativeainetworking,zhang2024generative}. Techniques such as meta-prompting and structured retrieval, continual learning, and parameter-efficient adaptation are measured by zero/low-shot performance, sample efficiency, adaptation time, and forgetting rate.
\end{itemize}

In the remainder of this survey, we operationalize these principles into a unifying taxonomy, a tutorial-style design flow with decision checklists, consolidated benchmarks and metrics for agentic networking, and reusable case-study templates that demonstrate how to instantiate compact, efficient, knowledgeable, and migratory agents in edge environments. The primary contributions of this survey are structured around addressing the existing gaps and challenges in the systematic exploration and deployment of Agentic AI tailored explicitly for edge general intelligence within 6G-enabled networks. Specifically, we aim to provide comprehensive insights into the conceptual distinctions, foundational design principles, practical deployment scenarios, and future research opportunities for Agentic AI and its agentification process, facilitating robust and scalable intelligent edge systems. The key contributions of this paper are summarized as follows:

\begin{itemize}
    \item We provide the first comprehensive survey and tutorial explicitly dedicated to multi-modal Agentic AI frameworks tailored for edge general intelligence within 6G-enabled networks. We clearly distinguish Agentic AI from traditional paradigms, including LLMs, MoEs, foundation models, and embodied AI frameworks, highlighting its unique characteristics and transformative potential.
    
    \item We systematically identify and elaborate on four foundational design pillars compactness, efficiency, migration, and knowledge \& reasoning that define the essential capabilities and requirements for practical, scalable, and explainable Agentic AI systems and agentification.
    
    \item We illustrate the transformative impact of Agentic AI through concrete use cases involving cooperative UAV swarms, adaptive vehicular networks, and edge robotics, emphasizing practical deployment scenarios and performance advantages.
    
    \item We analyze emerging open-source frameworks and critically discuss unresolved research challenges and promising future directions. These insights provide actionable guidance to enable robust, trustworthy, and scalable Agentic AI deployments across heterogeneous edge environments.
\end{itemize}

\begin{figure*}[t]
  \centering
  \includegraphics[width=0.9\linewidth]{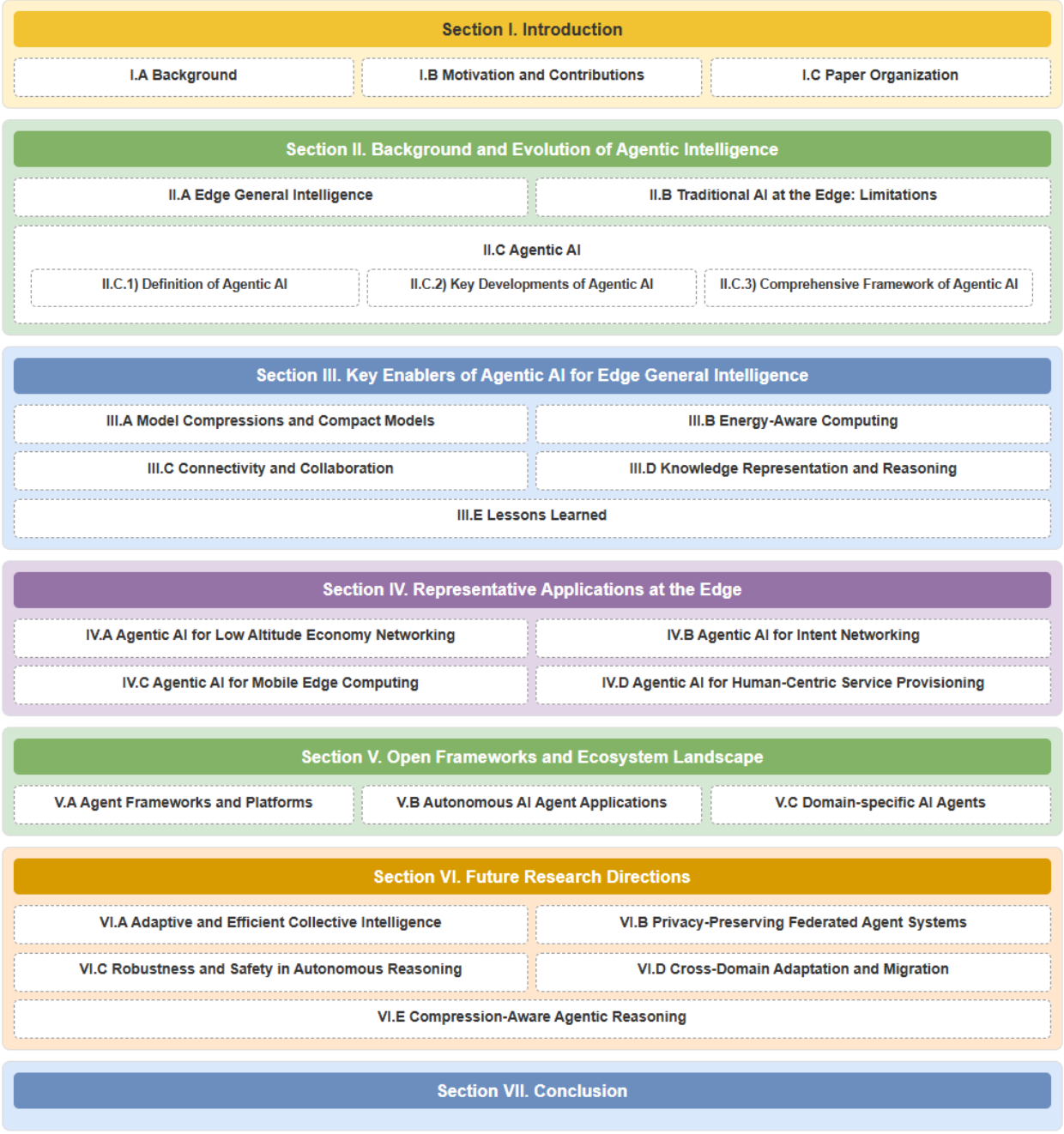}
  \caption{{Overall organization of this survey. We first introduces the evolution and core foundations of Agentic AI at the edge, followed by key enabling technologies and representative application scenarios. Subsequent sections address system design challenges, open-source frameworks, and future research directions, forming a coherent and layered roadmap toward edge general intelligence.}}
  \label{fig:architecture}
\end{figure*}

By engaging with this comprehensive survey, the readers will gain valuable insights into how to effectively adopt Agentic AI frameworks tailored to their specific edge intelligence applications. Additionally, the readers will deepen their understanding of practical deployment challenges, including joint offloading strategies for LLM-based models, dynamic model migration techniques across heterogeneous edge devices, and efficient routing methods for multi-LLM service orchestration. Such detailed knowledge will empower researchers and practitioners to better navigate the complexities and opportunities inherent to next-generation edge environments, ultimately fostering the development and realization of robust, scalable, and intelligent Agentic AI solutions and agentification process.

\subsection{Paper Organization}
As depicted in Fig.~\ref{fig:architecture}, the remainder of this paper is organized as follows. Section II introduces core concepts and frameworks of Agentic AI, emphasizing key capabilities and distinguishing it from traditional edge intelligence. Section III explores essential enabling technologies for Agentic AI and agentification process, including compact models, energy-aware computing, robust connectivity, and advanced reasoning techniques. Section IV presents representative Agentic AI applications in low-altitude economy, intent networking, vehicular networks, and human-centric service provisioning, supported by experimental analyses. Section V discusses critical system-level challenges and practical deployment strategies. Section VI reviews emerging open-source frameworks and toolkits. Section VII highlights promising future research directions, and Section VIII concludes the survey.


\section{Background}
\begin{table*}[!t]
\renewcommand\arraystretch{1.4}
\belowrulesep=0pt
\aboverulesep=0pt
\centering
\caption{Comparison of Edge General Intelligence and Traditional Edge Intelligence.}
\label{tab:egi_vs_ei}
\begin{tabular}{>{\centering\arraybackslash}m{0.15\textwidth}|| 
                >{\raggedright\arraybackslash}m{0.37\textwidth}| 
                >{\raggedright\arraybackslash}m{0.41\textwidth}}
\toprule
\midrule
\textbf{Feature} & \textbf{Edge General Intelligence} & \textbf{Traditional Edge Intelligence} \\
\midrule
\toprule
\cline{1-1}

Generalization 
& \textbf{Multi-task capability} \newline Supports multiple diverse tasks without retraining (vision, NLP, decision-making)
& \textbf{Task-specific models} \newline Designed specifically for single tasks (object detection, activity recognition) \\ \hline

Adaptability 
& \textbf{Dynamic adaptation} \newline Learns and adapts dynamically at runtime
& \textbf{Static behavior} \newline Requires manual retraining or updates \\ \hline

Model Architecture 
& \textbf{Compact general models} \newline Compact LLMs, mixture-of-experts (MoE), or foundation models optimized for edge \cite{du2024mixture}
& \textbf{Specialized models} \newline Small CNNs, RNNs, or DNNs optimized per task \\ \hline

Multi-Modality 
& \textbf{Multi-modal processing} \newline Handles text, images, audio, sensor fusion simultaneously
& \textbf{Single-modal processing} \newline Processes one modality at a time (e.g., images or sensors) \cite{borazjani2025multi} \\ \hline

Autonomy \& Reasoning 
& \textbf{Autonomous reasoning} \newline Independent decision-making with minimal cloud support
& \textbf{Inference-driven} \newline Executes predefined tasks with limited autonomy \\ \hline

Continual Learning 
& \textbf{Continuous learning} \newline Supports lifelong or federated learning directly on device
& \textbf{Limited online learning} \newline Rarely supports online learning due to resource constraints \\ \hline

Communication Dependency 
& \textbf{Low cloud dependency} \newline Reduced reliance on cloud, enhanced local processing \cite{zhao2022trine}
& \textbf{High cloud dependency} \newline Relies heavily on cloud for complex tasks \\ \hline

Personalization 
& \textbf{Dynamic personalization} \newline Automatically adjusts to user preferences \cite{chen2024towards}
& \textbf{Limited personalization} \newline Requires manual fine-tuning \\ \hline

Cognitive Collaboration 
& \textbf{Collaborative cognition} \newline Shares knowledge collaboratively with other agents
& \textbf{Isolated cognition} \newline Operates independently or strictly cloud-controlled \cite{alamouti2024building} \\ \hline

Security \& Privacy 
& \textbf{Enhanced local privacy} \newline Increased privacy via general-purpose on-device cognition
& \textbf{Cloud-dependent privacy} \newline Security contingent on data transmitted to cloud \\

\bottomrule
\end{tabular}
\end{table*}

\subsection{Edge General Intelligence}

Edge general intelligence represents an emerging paradigm aiming to extend generalized, adaptive, and context-aware cognitive capabilities directly onto resource-constrained edge devices~\cite{10876185,zhao2025worldmodelscognitiveagents}. Different from traditional edge intelligence, which primarily deploys task-specific, static models optimized for individual tasks (e.g., object detection or keyword spotting), edge general intelligence emphasizes versatility, adaptability, and autonomous cognitive reasoning~\cite{xu2021edge,xu2020edge}. Edge general intelligence leverages foundation models, such as compact LLMs, MoE, or multimodal neural architectures, to enable devices to perform multiple diverse tasks without frequent retraining, dynamically adapting to varying contexts, environments, and user preferences in real-time \cite{liu2024survey,xu2025decentralizationgenerativeaimixture}. In particular, reasoning capabilities, such as task decomposition, planning, and tool usage, are central to enabling goal-directed autonomy in dynamic edge environments. Such intelligent autonomy significantly reduces reliance on cloud-based resources, enhancing data privacy, operational efficiency, and user personalization~\cite{yang2025agentic,krishnan2025advancing}. Moreover, edge general intelligence systems increasingly integrate world-modeling capabilities, allowing agents to anticipate environmental dynamics and predict future states, which further strengthens their proactive planning and decision-making processes~\cite{gill2024edgeaitaxonomysystematic,su2025surveyautonomyinducedsecurityrisks}.

By employing continual learning strategies, cognitive collaboration between edge agents, and robust multimodal reasoning, edge general intelligence systems can continuously evolve and improve their cognitive capabilities throughout their deployment lifecycle \cite{10876185,alamouti2024building}. Typical application scenarios demonstrating the advantages of edge general intelligence include sophisticated smart home assistants capable of understanding complex user requests and contexts, and industrial IoT deployments that autonomously manage equipment maintenance, scheduling, and anomaly detection without extensive manual intervention or frequent model updates \cite{xu2021edge,zaidan2020review,dou2023artificialgeneralintelligenceagi}. To clearly illustrate the differences between edge general intelligence and traditional edge intelligence, Table~\ref{tab:egi_vs_ei} provides a comprehensive comparison across key technical dimensions.

\begin{figure*}[!t]
  \centering
  \includegraphics[width=0.95\linewidth]{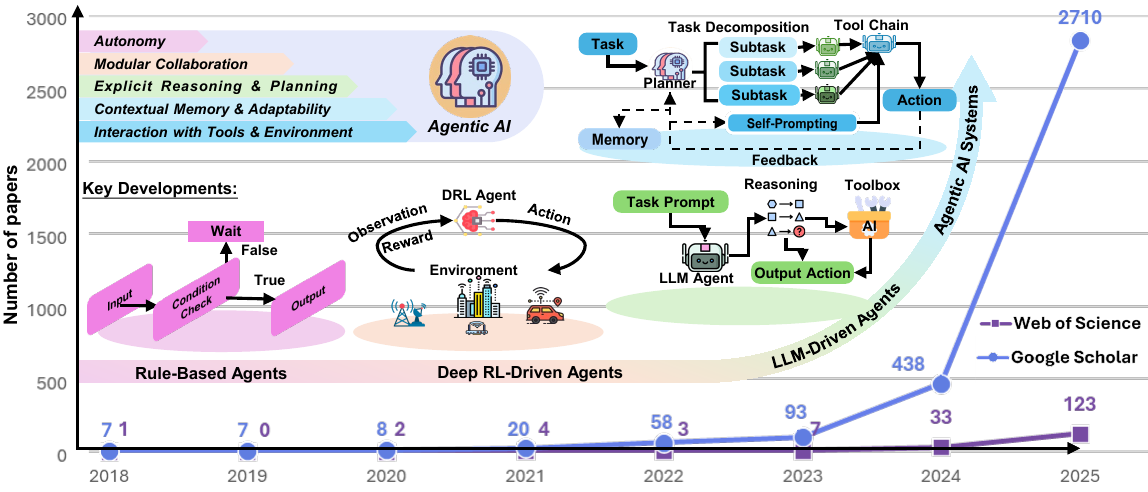}
  \caption{Illustration of key developments and evolution trajectory of Agentic AI systems, from early rule-based approaches, through DRL-driven agents, towards current LLM-driven agents. Highlighted are core capabilities such as autonomy, modular collaboration, explicit reasoning and planning, contextual memory and adaptability, and interaction with tools and environments. Recent trends emphasize task decomposition and self-prompting for robust reasoning and action execution, indicating substantial growth in research and deployment in the coming years.}
  \label{fig:key}
\end{figure*}

\subsection{Traditional AI at the Edge: Limitations}

Traditional edge AI systems have largely been designed for constrained, pre-defined tasks, such as object detection, speech recognition, or anomaly monitoring, operating under stable environments and with significant reliance on cloud infrastructure \cite{shi2020communication,9723563,10973078,9885226}. These designs, while effective in early passive scenarios, fall short in meeting the stringent requirements of emerging 6G networks and the broader vision of edge general intelligence~\cite{10081195,10258360,11066175}. In contrast to edge general intelligence, traditional edge intelligence lacks autonomy, adaptability, and context-aware reasoning capabilities essential for diverse, rapidly evolving operational environments \cite{azari2022evolution,8674240,8852687}. Specifically, the key limitations of traditional AI at the edge can be grouped into the following dimensions:

\begin{itemize}
    \item \textbf{Heavy reliance on cloud connectivity and centralized control:} Traditional approaches frequently depend on cloud infrastructure for model inference and training updates, introducing significant latency, bandwidth bottlenecks, and single points of failure \cite{eshratifar2019jointdnn,alieldin2021hiddencostedgeperformance,eugster2023toward}. This centralized design severely limits scalability and robustness, especially in decentralized and latency-sensitive edge scenarios.

    \item \textbf{Limited adaptability to dynamic environments:} Traditional edge AI models are typically static, lacking mechanisms for continuous adaptation to changing network conditions or user behaviors \cite{mcenroe2022survey,adhikari20226g,8766985}. Such static architectures face severe performance degradation in non-stationary edge environments, underscoring the need for more adaptive and continually evolving intelligent agents.

    \item \textbf{Scalability and real-time constraints under limited resources:} Conventional AI deployments at the edge often neglect tight resource constraints, such as limited memory, compute capacity, and power budgets, leading to inefficiencies in energy usage and operational responsiveness \cite{dev2025advanced,li2019edge}. To sustainably meet real-world demands of next-generation edge intelligence, future agents must integrate compact models, hardware-aware designs, and adaptive computational mechanisms.
\end{itemize}

These limitations highlight the need for a fundamental shift in edge intelligence, from cloud-reliant, static systems to intelligent agents that can operate autonomously and adaptively~\cite{10546122,bansod2025distinguishingautonomousaiagents}. This shift marks the evolution from the traditional {Internet of Things (IoT)}, where devices merely sense and transmit, to a more proactive {Internet of Agents (IoA)} powered by {Agentic AI}, where edge nodes perceive, reason, plan, and act independently in real time \cite{wang2025internet,9792233}. Next, we trace how AI agents have evolved toward this Agentic form.

\subsection{Agentic AI}

\subsubsection{Definition of Agentic AI}

Agentic AI refers to a new class of AI systems that exhibit goal-driven autonomy, operating in continuous \textit{perception-reasoning-action} loops \cite{acharya2025Agentic,miehling2025agenticaineedssystems}. Unlike conventional assistants that respond passively to user prompts, Agentic systems can proactively decompose high-level tasks, generate sub-goals, plan actions, and interact with external tools or environments with minimal human input \cite{sapkota2025aiagentsvsagentic,molinari2025pervasivedistributedagenticgenerative}. Powered by foundation models, these agents are designed to reason, act, and adapt over time. As described by IBM and Deloitte\footnote{\url{https://www.ibm.com/think/topics/Agentic-ai}}, Agentic AI systems are capable of completing complex workflows and achieving objectives with little or no human supervision~\cite{10962241}. Beyond individual autonomy, Agentic AI and agentification process often involve \textit{agent orchestration}, the coordinated interaction among multiple agents with specialized roles, enabling complex task execution through modular collaboration~\cite{tran2025multiagentsurvey,9171865,10818456}. Such orchestration allows agents to dynamically communicate, delegate subtasks, and synthesize partial outputs, forming a distributed problem-solving network especially suited for edge-centric, decentralized environments~\cite{li2025collaborativeinferencelearningedge}.

These Agentic AI systems are characterized by several core capabilities \cite{sapkota2025aiagentsvsagentic}. First, they exhibit {autonomy}, which enables decision making and action initiation, exemplified by UAVs navigating uncertain terrains or edge robots adapting to task variations. Second, they possess contextual memory and adaptability, allowing them to learn from past interactions and effectively respond to dynamic conditions such as those found in vehicular or industrial networks. Third, they support {explicit reasoning and planning}, utilizing external tools, APIs, or supplementary models for executing long-term strategies, as demonstrated in frameworks like ReAct~\cite{ReAct} and Toolformer~\cite{Toolformer}. Lastly, they facilitate {modular collaboration} by orchestrating toolchains across decentralized environments, supporting scalable deployment through platforms including HuggingGPT~\cite{HuggingGPT} and LangGraph\footnote{\url{https://github.com/langchain-ai/langgraph}.}~\cite{wang2025empiricalresearchutilizingllmbased}.

By design, Agentic AI addresses key limitations associated with traditional edge intelligence. It reduces dependency on cloud connectivity through localized inference and planning, enhances adaptability via memory-driven continuous learning, and improves computational efficiency by employing model compression and task-aware reasoning~\cite{jiang2025large,omidi2025memoryaugmentedtransformerssystematicreview}. These characteristics position Agentic AI as a foundational enabler for the {IoA}, converting passive devices into intelligent, self-directed agents capable of handling the complexities inherent to 6G networks and beyond~\cite{11103462,khowaja2025integrationagenticai6g}.

\begin{table*}[!t]
\centering
\footnotesize
\caption{Comparison of AI Agent Paradigms in Wireless Edge Intelligence.}
\label{tab:Agentic_comparison_transposed}
\begin{tabularx}{\textwidth}{m{2.8cm} m{3.2cm} m{3.2cm} m{3.2cm} m{3.2cm}}
\toprule
\textbf{Attribute} & \textbf{Rule-Based Agents} & \textbf{DRL-Driven Agents} & \textbf{LLM-Driven Agents} & \textbf{Agentic AI Systems} \\
\midrule

\textbf{Core Intelligence} & Finite-state machines, deterministic logic & DQN, PPO, A3C & Pre-trained transformers (GPT-2/3, Codex) & LLM + memory + DRL planner \\ \hline

\textbf{Autonomy} & {\color{red}\ding{55}} Manual, reactive &  Task-level &  Prompt-level autonomy & {\color{green}\checkmark} Goal-level autonomy \\ \hline 

\textbf{Perception Modality} & Single-modal (e.g., RSSI, CSI) & Env. features (QoS, SINR) & Text, code, limited images & Multi-modal (vision, RF, state/action) \\ \hline

\textbf{Memory Scope} & None (stateless) & Short-term (recurrent states) & Short-term buffer (few-shot) & Long-term (episodic vector DB) \\\hline

\textbf{Planning \& Reasoning} & None (fixed rules) & MDP-based finite-horizon policy & Prompt chaining, CoT-style reasoning & Deliberative planning, causal reasoning, recursive loops \\\hline

\textbf{Wireless Application} & CRN cooperative sensing, e.g., OR-rule CSS~\cite{gharib2019enhanced} & RIS-SWIPT beamforming via PPO~\cite{zhang2023energy} & LLM-assisted RAN control with MoE-PPO~\cite{zhang2024generative} & AutoGPT/Voyager for UAV control, RSMA spectrum negotiation~\cite{zhang2024generative,wang2023voyager} \\\hline

\textbf{Limitations} & No adaptation, static logic, poor scalability & Domain-specific, lacks abstraction or transferability & Limited memory, external tool dependence, task fragility & Higher cost, safety/policy alignment, runtime constraints \\

\bottomrule
\end{tabularx}
\label{table:1}
\end{table*}


\subsubsection{Key Developments of Agentic AI}

The emergence of Agentic AI and its agentification process are rooted in a multi-stage evolution, moving from simple automation toward increasingly autonomous, memory-enabled, and context-aware intelligent systems \cite{tong2025wirelessagent}. As shown in Fig.~\ref{fig:key}, this evolution progresses through several distinct stages: from basic \textit{rule-based agents} to adaptable \textit{DRL-driven agents}, further evolving into sophisticated \textit{LLM-driven agents}, and ultimately culminating in fully autonomous \textit{Agentic AI systems}. Each stage signifies enhanced cognitive capability and improved adaptability, addressing the growing complexity of wireless and edge environments~\cite{Ullah2025DRL,li2025llmguideddrlmultitierleo,Zhang_2024}.
\begin{itemize}
\item \textbf{Rule-Based Agents:}  
These early agents rely on predefined rules or finite-state machines, limiting their adaptability and autonomy. Such systems operate reactively and are primarily effective in static and narrowly defined scenarios, thus falling short in dynamic wireless environments~\cite{10937075}.

\item \textbf{Deep RL-Driven Agents:}  
Agents driven by deep reinforcement learning (DRL) enhance adaptability through trial-and-error interactions with the environment. However, their applicability remains constrained by task specificity, lacking broader generalization and explicit reasoning capabilities across diverse scenarios~\cite{9161406}.

\item \textbf{LLM-Driven Agents:}  
LLM-driven agents leverage large-scale language models such as GPT-2 and GPT-3 \cite{brown2020language} as cognitive cores, enabling general reasoning, multi-step planning, and sophisticated tool interactions. Frameworks including Codex \cite{chen2021evaluating}, ReAct \cite{yao2023react}, and Toolformer \cite{schick2023toolformer} exemplify this transition through structured tool use and reasoning chains. Specifically, in the wireless domain, Zhang \emph{et al.} \cite{zhang2024generative} proposed an LLM-based architecture integrating retrieval-augmented generation (RAG) and MoE enhanced PPO (MoE-PPO) for satellite-terrestrial integration. Their approach achieved 95.3\% retrieval accuracy and improved throughput by 42.6\% compared to traditional SDMA techniques, demonstrating the effectiveness of LLM-driven agents in optimizing both semantic inputs and physical-layer strategies.

\item \textbf{Agentic AI Systems:}  
Agentic AI with agentification process integrates autonomy, contextual memory, explicit reasoning, and modular collaboration into unified systems capable of long-term planning and proactive decision-making. Recent open-source frameworks such as AutoGPT, BabyAGI, and Voyager \cite{wang2023voyager} introduced sophisticated functionalities, including recursive task decomposition, adaptive self-prompting, and dynamic feedback mechanisms. In wireless network optimization, Zhang \emph{et al.} \cite{zhang2025agenticaigenerativeinformation} extended these principles to 6G networks, developing autonomous agents capable of generating optimal spectrum and energy policies through context-aware retrieval and goal-driven reasoning. This progression highlights a critical shift from passive assistant-like AI towards proactive, adaptive agents capable of effectively managing decentralized, real-time operational environments.

\end{itemize}

To highlight the evolving design philosophies of AI agents in wireless systems, Table~\ref{tab:Agentic_comparison_transposed} compares four major paradigms, i.e., rule-based agents, deep RL-driven agents, LLM-driven agents, and Agentic AI, across key technical dimensions including core intelligence, autonomy, perception modality, memory scope, planning and reasoning capabilities, representative wireless applications, and known limitations.

\begin{figure*}[!t]
  \centering
  \includegraphics[width=\linewidth]{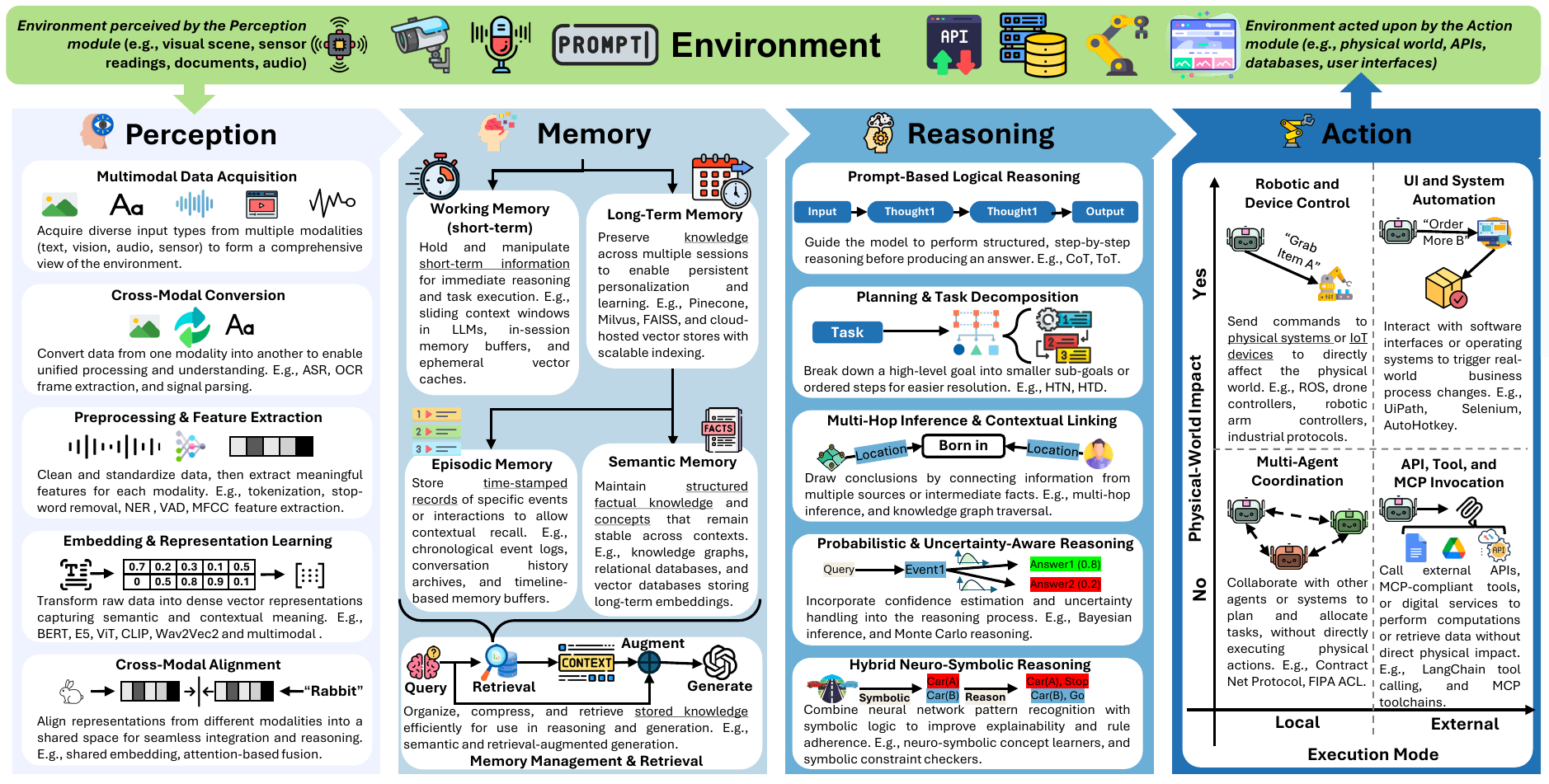}
   \caption{{Comprehensive workflow of Agentic AI for edge deployments. The pipeline comprises four modules: \emph{Perception} (i.e., multi-modal acquisition, cross-modal conversion, preprocessing, feature extraction, and embedding for a unified scene); \emph{Memory} (i.e., working, episodic, semantic, and long-term stores with management and retrieval supporting vector caches and RAG); \emph{Reasoning} (i.e., prompt-based logic, planning and decomposition, multi-hop/context linking, uncertainty-aware and neuro-symbolic inference); and \emph{Action} (i.e., robot or device control, multi-agent coordination, API or tool or MCP invocation, and UI or system automation)~\cite{li2025perceptionreasonthinkplan,salimpour2025embodiedagenticaireview}. Execution may be local on device or external at edge or cloud, with environmental feedback closing the loop. The figure also indicates where the four design principles apply: compactness, efficiency, knowledge and reasoning, and migration.}}
  \label{fig:flow}
\end{figure*}

\subsection{Comprehensive Framework of Agentic AI}

{Beyond understanding its foundational definition, it is essential to conceptualize Agentic AI through its comprehensive architecture, core capabilities, and primary functional components~\cite{ren2025aiagentsagenticainavigating}. As illustrated in Fig.~\ref{fig:flow}, a complete Agentic AI framework integrates several interconnected modules, enabling autonomous perception, reasoning, planning, and effective action execution. {Concretely, the architecture is organized into four modules, i.e., Perception, Memory, Reasoning, and Action, with an explicit memory management and retrieval layer that binds them together, and with execution either on-device or offloaded to the edge/cloud~\cite{11030757,mei2025surveycontextengineeringlarge}.} Specifically, the Agentic AI with agentification process operates through a continuous cycle, beginning from external data and environmental perception, moving through comprehension and reasoning stages, and culminating in adaptive actions that are continuously refined through feedback loops~\cite{liu2025advances,11071266,wu2025positionpaperopencomplex}.  The core components of the Agentic AI framework are outlined below, supported by concrete examples and recent research insights:}

\begin{itemize}

    \item \textbf{Perception Module:}  
    This module integrates multi-modal data, including textual, visual, and auditory inputs, allowing agents to perceive and understand complex environments comprehensively. For instance, autonomous vehicles leverage multi-modal sensor fusion, combining LiDAR, radar, camera images, and traffic signals, to accurately interpret real-time road conditions, pedestrian behaviors, and traffic patterns, thereby ensuring reliable and context-aware navigation~\cite{acharya2025Agentic,sapkota2025aiagentsvsagentic}.
    
    \item \textbf{LLMs:}  
    LLMs such as GPT-4 and Gemini function as cognitive cores, delivering rich semantic comprehension and sophisticated reasoning abilities. These capabilities enable agents to interpret complex instructions, decompose high-level tasks, and generate structured plans. For example, AutoGPT utilizes GPT-4 to autonomously interpret high-level commands, decomposing them into detailed subtasks, such as generating business strategies or optimizing complex workflows, thereby significantly reducing human supervision and enhancing operational efficiency~\cite{Significant_Gravitas_AutoGPT,chen2021evaluating}.
    
    \item \textbf{External Tools and APIs:}  
    Agentic AI seamlessly interact with external tools and APIs to perform actions that extend beyond their inherent cognitive capabilities. For example, Toolformer integrates API calls directly within the reasoning process, allowing agents to dynamically access external computational resources, such as mathematical computation APIs, databases, and specialized knowledge repositories, thereby facilitating complex problem-solving in real-time scenarios~\cite{schick2023toolformer,yao2023react}.
    
    \item \textbf{Memory and Retrieval:}  
    Memory components, particularly RAG mechanisms, enable agents to continuously learn and retain historical knowledge effectively~\cite{liu2025wirelessagenticairetrievalaugmented}. For instance, recent research by Wang et al.~\cite{wang2023augmenting} developed memory-augmented neural networks utilizing RAG techniques to dynamically retrieve contextually relevant information from vectorized knowledge bases. This approach significantly enhanced model adaptability and reasoning accuracy across diverse tasks, effectively supporting agentic decision-making in complex scenarios.

    \item \textbf{Planning and Reasoning:}  
    Explicit planning capabilities, including Chain-of-Thought (CoT)~\cite{wang2025chainofthoughtlargelanguagemodelempowered,zhang2024chain} reasoning and symbolic AI techniques, empower agents to formulate and assess long-horizon strategies autonomously. A notable example is the ReAct framework~\cite{yao2023react}, which combines language-driven reasoning with action planning. By integrating CoT-based reasoning methods, ReAct allows agents to systematically break down complex goals into manageable subtasks, dynamically evaluating potential outcomes, and selecting optimal actions, greatly enhancing their ability to manage sophisticated, real-time decision-making scenarios.
    
    \item \textbf{Multi-Agent Coordination:}  
    Multi-agent frameworks leveraging Deep DRL facilitate decentralized coordination, collaborative decision-making, and emergent collective intelligence. For instance, Tong et al.~\cite{tong2025wirelessagent} introduced a Multi-Agent DRL system incorporating model context protocol (MCP)~\cite{hou2025model} to enhance decentralized spectrum management in wireless networks. MCP supports efficient inter-agent communication by maintaining consistent context representations across distributed agents, significantly improving network throughput, reducing latency, and demonstrating the robustness and scalability of decentralized Agentic AI coordination.
    
\end{itemize}

\begin{figure}[!t]
  \centering
  \includegraphics[width=0.95 \linewidth]{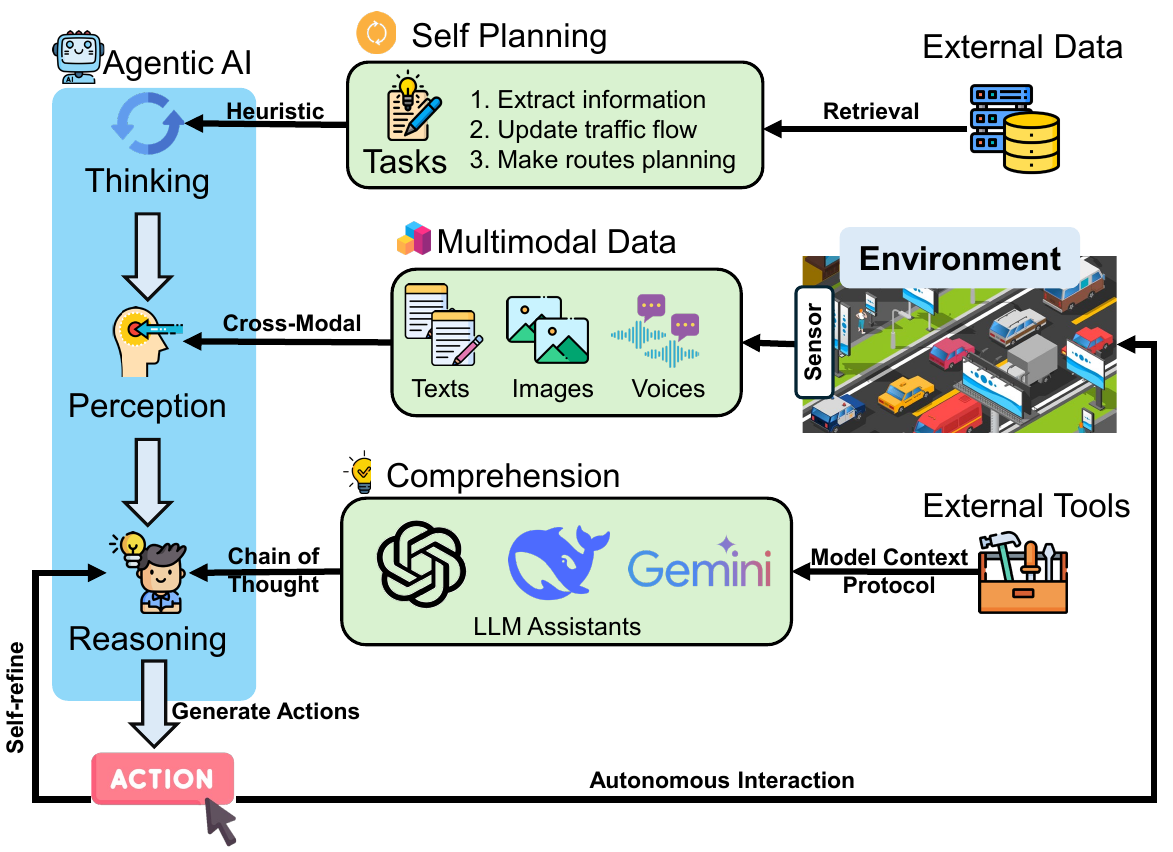}
  \caption{Conceptual workflow illustrating how Agentic AI autonomously integrates self-planning, multimodal perception, comprehension via foundation models, and external tools for continuous perception–reasoning–action agentification process in edge general intelligence systems~\cite{zhang2025agenticaigenerativeinformation}.}
  \label{fig:enabler_v1}
\end{figure}

Specifically, as shown in Fig.~\ref{fig:enabler_v1}~\cite{zhang2025agenticaigenerativeinformation}, these components collaborate within an integrated and iterative workflow as follows: Initially, the Perception Module captures external multimodal data and environmental context. Next, the Comprehension stage, driven by foundation models and LLMs, processes and interprets this data, providing rich semantic understanding and structured contextual insights. These insights inform the Self-Planning stage, wherein agents autonomously formulate tasks, strategies, and action plans. In the subsequent Reasoning stage, explicit planning mechanisms (e.g., CoT reasoning) and long-term memory retrieval (e.g., RAG) refine these plans further, incorporating additional contextual insights and prior knowledge. Finally, autonomous actions are executed using External Tools and APIs, with outcomes continuously evaluated and fed back into the loop for iterative self-refinement.

By integrating these components within this coherent and iterative perception-reasoning-action agentification process, Agentic AI establishes a robust cognitive architecture capable of autonomously adapting to dynamic edge environments~\cite{dev2025advanced,jiang2025large,fang2025comprehensivesurveyselfevolvingai}. This comprehensive integration addresses fundamental limitations of traditional edge intelligence, significantly advancing the realization of edge general intelligence and paving the way toward resilient and adaptive intelligent systems for next-generation edge deployments~\cite{10133894,li2025resilientfederatedlearningcyberedge,liu2025integrated}.

\section{Key Enablers of Agentic AI for Edge General Intelligence}
\begin{figure*}[!t]
  \centering
  \includegraphics[width=\linewidth]{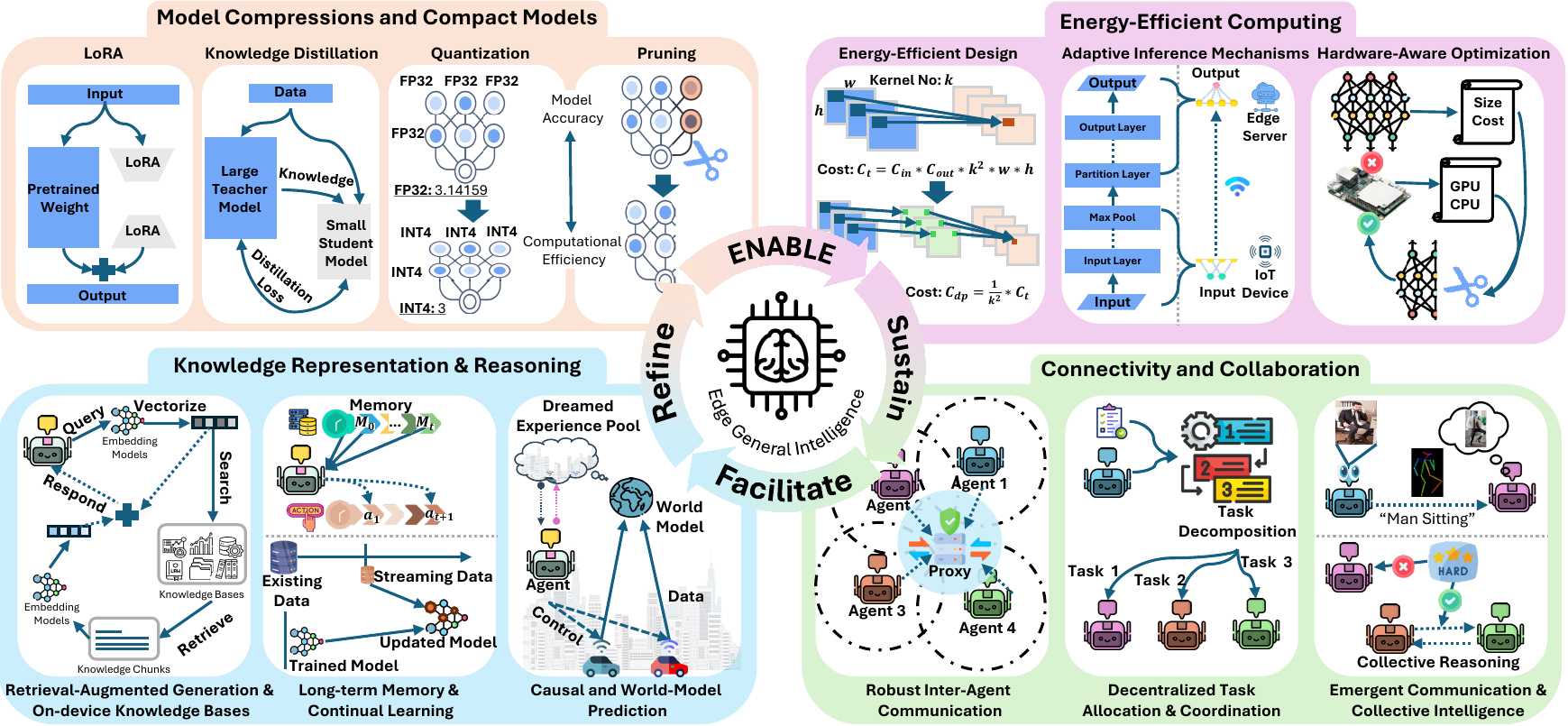}
  \caption{Interdependencies Among Key Enablers of Agentic AI for edge general intelligence. Compact model techniques enable efficient execution under tight resource constraints, thereby sustaining energy-aware computing. In turn, energy efficiency sustains continuous operation and supports robust connectivity and collaboration. These collaborative mechanisms facilitate distributed knowledge representation and reasoning. Finally, advanced reasoning capabilities guide the design of increasingly compact and adaptive models, forming a virtuous cycle that drives scalable and autonomous edge intelligence.}
  \label{fig:enabler}
\end{figure*}

To enable Agentic AI in the edge general intelligence, several key technological enablers must be addressed, including compact model deployment, energy-efficient computing, robust connectivity and collaboration, and effective knowledge representation and reasoning~\cite{9606720,huang2024digital}. These enablers collectively support edge general intelligence, ensuring agents can autonomously adapt and reason effectively in dynamic and resource-constrained environments~\cite{10811956,shen2025revolutionizing}.

\subsection{Model Compressions and Compact Models}

Deploying Agentic AI models directly onto resource-constrained edge devices presents significant challenges due to stringent memory and computational constraints~\cite{wang2025empowering}. To address these limitations while preserving critical reasoning and adaptability features inherent in Agentic AI, model-compression techniques such as pruning, quantization, low-rank factorization, and knowledge distillation are indispensable~\cite{10597596,1011453729218}. These compact model strategies are particularly crucial for supporting the {perception module} of Agentic AI and the deployment of sophisticated {foundation models and LLMs} in edge environments, allowing effective multimodal data interpretation and complex reasoning capabilities without exceeding computational budgets. Moreover, architectures explicitly designed for edge deployment, including MobileNets and ShuffleNet, employ depthwise separable convolutions, significantly minimizing computational load and latency~\cite{shahriar2025comparativeanalysislightweightdeep,asare2025deployingevaluatingmultipledeep}. These inherently efficient designs thus directly facilitate rapid and autonomous decision-making processes fundamental to Agentic AI, paving the way towards practical edge general intelligence~\cite{10849561,10857320,wu2022ai}.

\textit{LoRA:} Low-Rank Adaptation (LoRA) reduces the complexity of adapting large, pre-trained models by inserting small, trainable matrices into frozen weights. For example, Hu et al.~\cite{hu2022lora} proposed LoRA, demonstrating that inserting low-rank decompositions into Transformer layers can significantly decrease the number of trainable parameters by up to 10,000 times, without compromising performance. Specifically, experiments on GPT-3 (175B parameters) showed that LoRA achieved competitive or superior performance compared to full fine-tuning, while also reducing GPU memory consumption by approximately threefold. By significantly compressing these models, LoRA directly supports the efficient integration of powerful {LLMs} within Agentic AI systems, facilitating advanced semantic comprehension and reasoning on resource-limited edge hardware.

\textit{Knowledge Distillation:} Knowledge distillation further supports Agentic AI by transferring intricate reasoning capabilities, such as chain-of-thought reasoning, from large ``teacher'' models to compact ``student'' models optimized for edge deployment. For instance, Li et al.~\cite{li2023prompt} introduced a prompt-based distillation approach specifically targeting complex reasoning behaviors in LLMs. Their methodology effectively distilled multi-step reasoning capabilities into smaller models such as Llama2 (7B parameters) and CodeLlama, achieving accuracies of 85\% and 85.5\%, respectively, on the challenging SVAMP arithmetic reasoning dataset. These results surpassed GPT-3.5-turbo, demonstrating that complex cognitive reasoning can be effectively condensed into compact architectures suitable for edge environments. This approach is particularly relevant for enhancing the efficiency of the {planning and reasoning} module within Agentic AI, enabling high-level reasoning and explicit planning capabilities in constrained edge scenarios.

\textit{Quantization Methods:} Quantization methods, such as aggressive post-training quantization, reduce model parameters and activations to lower bit-precisions (e.g., 8-bit or 4-bit), enabling large-scale models to run efficiently on edge devices. Lin et al.~\cite{lin2024awq} proposed Activation-aware Weight Quantization (AWQ), specifically designed for quantizing LLMs with minimal performance degradation. They demonstrated lossless performance across 11 vision-language benchmarks, with INT4-g128 quantization settings applied to models such as VILA-7B and VILA-13B fully matching their original full-precision counterparts. These quantization techniques are crucial for enabling efficient real-time multimodal perception within the Agentic AI {perception module} and comprehensive interpretation within {LLMs}, maintaining the rich cognitive functionalities essential to Agentic AI deployments in resource-constrained environments.

\textit{Pruning Methods:} Structured pruning techniques such as SparseGPT and LLM-Pruner systematically remove redundant neurons, channels, or attention heads, creating efficient models while maintaining their functional integrity and reasoning performance. Ma et al.~\cite{ma2023llm} introduced LLM-Pruner, the first structured pruning framework explicitly designed for LLMs. Using only 50,000 training samples and three hours of fine-tuning, they achieved parameter reduction of up to 20\% while preserving over 94\% of the original model's performance. Pruning methods play a critical role in supporting the efficient integration of comprehensive reasoning, {memory and retrieval} mechanisms (e.g., vectorized databases and RAG), and autonomous {multi-agent coordination}, enabling compact yet highly capable Agentic AI models that thrive in decentralized and collaborative edge deployments.

\begin{table*}[ht]
\renewcommand\arraystretch{1.4}
\belowrulesep=0pt
\aboverulesep=0pt
\centering
\caption{Model Compression Techniques for edge general intelligence.}
\label{tab:model_compression_egi}
\begin{tabular}{
>{\centering\arraybackslash}m{0.13\textwidth}||
>{\raggedright\arraybackslash}m{0.03\textwidth}|
>{\raggedright\arraybackslash}m{0.18\textwidth}|
>{\raggedright\arraybackslash}m{0.18\textwidth}|
>{\raggedright\arraybackslash}m{0.18\textwidth}}
\toprule
\midrule
\rowcolor[gray]{0.9}
{\textbf{Technique}} &
{\textbf{Ref}} &
{\textbf{Description}} &
{\textbf{Pros}} &
{\textbf{Cons}} \\ 
\midrule
\toprule

LoRA & \cite{hu2022lora} 
& Inserts small trainable matrices into large frozen models 
& \textbullet~Major parameter reduction\newline\textbullet~Lower memory usage\newline\textbullet~Good performance
& \textbullet~Insertion complexity\newline\textbullet~Accuracy sensitive to rank \\ 
\midrule

Knowledge Distillation & \cite{li2023prompt}
& Transfers reasoning from large teacher to compact student models
& \textbullet~Preserves complex reasoning\newline\textbullet~Improves small-model performance
& \textbullet~Depends on teacher quality\newline\textbullet~Possible knowledge loss \\ 
\midrule

Quantization Methods & \cite{lin2024awq}
& Reduces parameter precision (e.g., INT4) to optimize efficiency
& \textbullet~Minimal accuracy loss\newline\textbullet~High edge efficiency
& \textbullet~Sensitive at ultra-low precision\newline\textbullet~Needs calibration \\ 
\midrule

Pruning Methods & \cite{ma2023llm}
& Removes redundant neurons or attention heads structurally
& \textbullet~Effective sparsity (up to 20\%)\newline\textbullet~Maintains high accuracy
& \textbullet~Performance loss at high sparsity\newline\textbullet~Pruning strategy required \\ 

\bottomrule
\end{tabular}
\end{table*}


\subsection{Energy-Aware Computing}

The deployment of Agentic AI on resource-constrained edge devices inherently demands energy-aware computing strategies~\cite{9865979}. Agentic AI models, characterized by autonomous reasoning, proactive decision-making, and long-term operational autonomy, necessitate continuous yet efficient execution under limited power budgets and strict thermal constraints~\cite{yang2020edgeintelligenceautonomousdriving}. Energy-aware computing methods thus become fundamental to enabling these intelligent agents to perform sophisticated reasoning tasks reliably and sustainably at the edge, directly advancing edge general intelligence.

\textit{Energy-Efficient Model Design:} 
Efficient execution of Agentic AI on edge hardware mandates model architectures specifically designed to minimize computational and energy footprints. Compact model architectures such as MobileNet~\cite{luo2020comparison}, ShuffleNet~\cite{zhang2018shufflenet}, and EfficientNet~\cite{tan2019efficientnet} leverage depthwise separable convolutions, channel shuffle operations, and compound scaling strategies, significantly reducing energy consumption and latency without compromising reasoning performance. Recent advancements extend these concepts to Transformer-based architectures optimized for energy efficiency. For example, MobileViT~\cite{mehta2021mobilevit} effectively integrates vision transformers into edge-friendly models, enabling sophisticated visual reasoning at minimal energy cost, thus driving the practical realization of edge general intelligence. 

\textit{Adaptive Inference Mechanisms:} 
Agentic AI benefits profoundly from adaptive computation strategies that dynamically adjust computational resources in response to input complexity and environmental constraints. Techniques such as dynamic neural networks~\cite{han2021dynamic} and multi-exit architectures~\cite{bakhtiarnia2021multi, addad2023multi} enable conditional execution of neural pathways or early termination of inference based on confidence levels, significantly reducing redundant computations. For instance, recent work demonstrates adaptive early-exit schemes achieving up to 24.6\% latency and 46.5\% energy consumption reductions compared to state-of-the-art IoT ML inference, adaptively distributing computation between devices and edge servers without accuracy loss~\cite{9700550}. Such adaptive mechanisms are particularly vital for enhancing the efficiency of the {planning and reasoning} processes within Agentic AI, ensuring that high-level, explicit reasoning and complex decision-making tasks can be executed within real-time constraints of edge environments.

\textit{Hardware-Aware Optimization:}
Effective integration of Agentic AI into edge general intelligence requires optimized software-hardware co-design, aligning computational tasks explicitly with the capabilities of specialized edge accelerators (e.g., NPUs, TPUs, and VPUs). Techniques such as hardware-aware neural architecture search (HW-NAS)~\cite{benmeziane2021comprehensive, li2021hw} and targeted model pruning~\cite{dave2021hardware} adapt neural network structures to specific accelerator architectures, exploiting hardware strengths and minimizing costly memory transfers and computations. Additionally, dynamic voltage and frequency scaling (DVFS) coupled with flexible invocation-based deep reinforcement learning~\cite{10689358} enables flexible adjustment of agent invocation intervals and task scheduling, achieving a 55.1\% reduction in agent invocation cost and up to 23.3\% overall energy consumption reduction. These hardware-aware optimizations are essential for efficiently running computationally intensive components such as {memory and retrieval (e.g., RAG)}, the execution of {external tools and APIs}, and supporting robust decentralized operations essential for {multi-agent coordination}, collectively contributing to the energy-efficient and scalable deployment of Agentic AI in edge general intelligence contexts.

\subsection{Connectivity and Collaboration}

Effective collaboration and robust connectivity are foundational to Agentic AI systems, enabling decentralized agents to seamlessly cooperate, share intelligence, and execute complex tasks at the edge~\cite{8863721,ranjan2025lokaprotocoldecentralizedframework}. Given the inherently distributed and dynamic nature of edge general intelligence, robust and efficient communication protocols, collaborative decision-making algorithms, and adaptive coordination strategies become critical for the scalable and reliable deployment of intelligent agents in real-world environments~\cite{10546304,jin2025comprehensivesurveymultiagentcooperative}.

\begin{table*}[ht]
\renewcommand\arraystretch{1.4}
\belowrulesep=0pt
\aboverulesep=0pt
\centering
\caption{Connectivity and Collaboration Techniques for Edge General Intelligence}
\label{tab:connectivity_collaboration_egi}
\begin{tabular}{
>{\centering\arraybackslash}m{0.14\textwidth}||
>{\raggedright\arraybackslash}m{0.08\textwidth}|
>{\raggedright\arraybackslash}m{0.20\textwidth}|
>{\raggedright\arraybackslash}m{0.20\textwidth}|
>{\raggedright\arraybackslash}m{0.20\textwidth}}
\toprule
\midrule
\rowcolor[gray]{0.9}
{\textbf{Technique}} &
{\textbf{Ref}} &
{\textbf{Description}} &
{\textbf{Pros}} &
{\textbf{Cons}} \\ 
\midrule
\toprule

Robust Inter-Agent Communication 
& \cite{kempe2003gossip, mcmahan2017communication, dai2017improved}
& Gossip algorithms, federated learning, and sparse message passing for robust and efficient communication
& \textbullet~Resilience to failures\newline\textbullet~Low bandwidth usage\newline\textbullet~Efficient dissemination
& \textbullet~Message latency\newline\textbullet~Potential slow convergence\\ 
\midrule

Decentralized Task Allocation and Coordination 
& \cite{foerster2016learning, lowe2019pitfalls, fioretto2018distributed, li2020deep}
& Multi-agent RL, distributed constraint optimization, and graph neural networks for decentralized decision-making
& \textbullet~Dynamic adaptability\newline\textbullet~Scalable coordination\newline\textbullet~No central control required
& \textbullet~Training complexity\newline\textbullet~Coordination difficulty\\ 
\midrule

Emergent Communication and Collective Intelligence 
& \cite{foerster2016learning, lazaridou2020emergent}
& MARL-driven autonomous development of concise and adaptive communication protocols
& \textbullet~Reduced overhead\newline\textbullet~Efficient semantics\newline\textbullet~Enhanced scalability
& \textbullet~Complex training\newline\textbullet~Difficult interpretability\\ 

\bottomrule
\end{tabular}
\end{table*}

\textit{Robust Inter-Agent Communication:}
Reliable communication under intermittently connected or bandwidth-constrained edge scenarios is crucial for coordinating actions among distributed agents. Recent studies have focused on low-overhead, resilient communication protocols such as gossip-based algorithms~\cite{kempe2003gossip}, federated learning protocols~\cite{mcmahan2017communication}, and sparse message-passing schemes~\cite{dai2017improved}, which effectively propagate information through the network with minimal redundancy. These methods enhance the robustness of Agentic AI systems against link failures, bandwidth fluctuations, and intermittent connectivity, thus maintaining collective intelligence and enabling seamless collaboration in edge general intelligence frameworks. Such robust communication protocols directly support the efficient integration and coordination of the {perception module} and the effective exchange of multimodal data processed by {LLMs}, ensuring reliable information sharing across distributed agent networks under challenging conditions~\cite{kong2025surveyllmdrivenaiagent}.

\textit{Decentralized Task Allocation and Coordination:}
Agentic AI deployed in edge environments necessitates decentralized and self-organizing mechanisms for task distribution and resource allocation. Techniques from multi-agent reinforcement learning~\cite{foerster2016learning,lowe2019pitfalls,9195488}, distributed constraint optimization~\cite{fioretto2018distributed}, and graph neural network-based coordination methods~\cite{li2020deep} have been successfully leveraged for decentralized decision-making. For instance, graph-based coordination frameworks allow distributed agents to perform collaborative inference and task allocation without centralized control, dynamically adapting to environmental changes and agent availability, thus significantly advancing the autonomy and scalability of edge general intelligence~\cite{10856273}. These decentralized coordination techniques directly enable effective {planning and reasoning} among distributed agents, while also optimizing the invocation of {external tools and APIs}, allowing for efficient resource usage and improved collective performance in dynamic edge scenarios.

\textit{Emergent Communication and Collective Intelligence:}
Enabling Agentic AI systems to autonomously develop efficient, concise, and adaptive communication languages or signaling mechanisms greatly enhances their collaborative capabilities at the edge. Recent research on emergent communication~\cite{foerster2016learning,lazaridou2020emergent} demonstrates that agents can autonomously learn shared languages optimized for minimal communication overhead while efficiently encoding task-relevant semantics. To design such systems, multi-agent reinforcement learning (MARL) frameworks, such as those demonstrated in~\cite{foerster2016learning}, can be employed to train agents to optimize communication protocols, which involves using discrete message spaces to ensure conciseness and defining reward structures that balance task performance with communication efficiency, such as rewarding task completion while penalizing excessive messaging. Additionally, integrating lightweight attention-based modules ensures efficient communication on resource-constrained edge devices. Such emergent collective intelligence paradigms allow edge-deployed agent groups to perform complex tasks collaboratively with minimal communication resources, significantly reducing energy usage and enhancing scalability, thereby promoting robust and adaptive edge general intelligence~\cite{xu2021edge}. Furthermore, emergent communication enhances the functionality of the {memory and retrieval (e.g., RAG)} mechanisms, allowing agents to effectively encode and recall shared experiences. This facilitates seamless collaboration and supports advanced decentralized operations critical for scalable and efficient {multi-agent coordination}~\cite{10620276}.

\subsection{Knowledge Representation \& Reasoning}

Effective knowledge representation and reasoning capabilities are foundational to Agentic AI, enabling intelligent agents to anticipate future states, reason about their environment, and continually adapt through learning. In the context of edge general intelligence, these capabilities must be realized efficiently and robustly within resource-constrained environments. Key techniques include Retrieval-Augmented Generation (RAG), on-device knowledge bases, long-term memory integration, causal and world-model predictions, and continual learning.

\textit{Retrieval-Augmented Generation (RAG) and On-device Knowledge Bases:}
RAG significantly enhances agent capabilities by integrating external knowledge bases during inference, improving reasoning accuracy and factual consistency~\cite{10531073}. Recent advancements in compact vector databases and efficient retrieval algorithms~\cite{han2023comprehensive} enable on-device storage and rapid retrieval of relevant information with minimal computational overhead. For example, lightweight retrieval systems allow edge-deployed language models to dynamically access and utilize up-to-date external data locally without constant external connectivity. 

\textit{Long-term Memory and Continual Learning:}
Agentic AI necessitates mechanisms for retaining and reasoning over extended temporal contexts, continuously updating internal knowledge representations. Long-term memory architectures such as memory-augmented neural networks~\cite{wang2023augmenting} and transformer models with extended memory modules~\cite{rae2020transformers} efficiently store and retrieve historical knowledge. Additionally, lightweight continual-learning frameworks~\cite{paissan2024structured,li2019learn} allow edge agents to incrementally assimilate new information without catastrophic forgetting, significantly enhancing adaptability and operational autonomy. 

\textit{Causal and World-Model Prediction:}
Causal reasoning and world-model prediction capabilities enable Agentic AI systems to understand environmental dynamics, anticipate outcomes, and proactively perform look-ahead planning entirely on edge devices. Techniques such as latent dynamics modeling~\cite{hafner2019learning}, causal reinforcement learning~\cite{zhang2020designing}, and predictive simulation frameworks~\cite{cullen2023predicting} offer computationally efficient models of environmental interactions. World models, in particular, enable agents to internally simulate future states, evaluate potential actions, and select optimal strategies without expensive real-world trial-and-error interactions. This capability significantly enhances sample efficiency, safety, and planning effectiveness. 

\subsection{Lessons Learned}

The deployment of Agentic AI on resource-constrained edge devices requires integrated solutions across multiple technical fronts. Compact model techniques, such as low-rank adaptation, quantization, pruning, and distillation, enable efficient execution under strict memory and compute budgets~\cite{han2023comprehensive}. Energy-aware architectures and adaptive inference, combined with hardware-level optimizations, ensure sustainable operation within power and thermal limits~\cite{luo2020comparison}. Robust communication and decentralized coordination strategies allow agents to collaborate effectively in dynamic environments~\cite{qiu2024collaborative}. Morevoer, advanced knowledge representation methods such as retrieval-augmented generation, long-term memory, and causal reasoning, support adaptive decision-making and future state prediction~\cite{10531073}. Together, these capabilities form the foundation for scalable, autonomous, and intelligent Agentic AI at the edge.

\begin{table*}[ht]
\renewcommand\arraystretch{1.4}
\belowrulesep=0pt
\aboverulesep=0pt
\centering
\caption{Energy-Aware Computing Techniques for Edge General Intelligence}
\label{tab:energy_aware_egi}
\begin{tabular}{
>{\centering\arraybackslash}m{0.14\textwidth}||
>{\raggedright\arraybackslash}m{0.08\textwidth}|
>{\raggedright\arraybackslash}m{0.19\textwidth}|
>{\raggedright\arraybackslash}m{0.19\textwidth}|
>{\raggedright\arraybackslash}m{0.18\textwidth}}
\toprule
\midrule
\rowcolor[gray]{0.9}
{\textbf{Technique}} &
{\textbf{Ref}} &
{\textbf{Description}} &
{\textbf{Pros}} &
{\textbf{Cons}} \\ 
\midrule
\toprule

Energy-Efficient Model Design & \cite{luo2020comparison, zhang2018shufflenet, tan2019efficientnet, mehta2021mobilevit} 
& Designs compact models (e.g., MobileNet, ShuffleNet, EfficientNet, MobileViT) optimized for energy efficiency at the edge
& \textbullet~Reduced computational cost\newline\textbullet~Minimal energy footprint\newline\textbullet~High visual reasoning capability
& \textbullet~Potential accuracy-performance trade-off\newline\textbullet~Limited capacity in complex tasks\\ 
\midrule

Adaptive Inference Mechanisms & \cite{han2021dynamic, bakhtiarnia2021multi, addad2023multi,9700550}
& Dynamically adjusts computational resources based on input complexity, using dynamic neural networks and multi-exit architectures
& \textbullet~Significant latency (24.6\%) and energy (46.5\%) reductions\newline\textbullet~Adaptive computation without accuracy loss
& \textbullet~Increased design complexity\newline\textbullet~Possible miscalibration at inference\\ 
\midrule

Hardware-Aware Optimization & \cite{benmeziane2021comprehensive, li2021hw, dave2021hardware,10689358}
& Co-designs models and hardware via hardware-aware NAS, targeted pruning, and DVFS with DRL-based invocation scheduling~\cite{11059888}
& \textbullet~Reduced energy (up to 23.3\%) and invocation cost (55.1\%)\newline\textbullet~Optimized alignment with edge accelerators (NPUs, TPUs, VPUs)
& \textbullet~Hardware-specific optimization overhead\newline\textbullet~Limited portability across hardware platforms\\ 

\bottomrule
\end{tabular}
\end{table*}


\begin{table*}[ht]
\renewcommand\arraystretch{1.4}
\belowrulesep=0pt
\aboverulesep=0pt
\centering
\caption{Knowledge Representation and Reasoning Techniques for Edge General Intelligence}
\label{tab:knowledge_reasoning_egi}
\begin{tabular}{
>{\centering\arraybackslash}m{0.15\textwidth}||
>{\raggedright\arraybackslash}m{0.08\textwidth}|
>{\raggedright\arraybackslash}m{0.20\textwidth}|
>{\raggedright\arraybackslash}m{0.21\textwidth}|
>{\raggedright\arraybackslash}m{0.20\textwidth}}
\toprule
\midrule
\rowcolor[gray]{0.9}
{\textbf{Technique}} &
{\textbf{Ref}} &
{\textbf{Description}} &
{\textbf{Pros}} &
{\textbf{Cons}} \\ 
\midrule
\toprule
RAG and On-device Knowledge Bases 
& \cite{han2023comprehensive}
& Integrates compact vector databases and efficient retrieval algorithms for on-device dynamic knowledge access
& \textbullet~High factual consistency\newline\textbullet~Robust offline autonomy\newline\textbullet~Low computational overhead
& \textbullet~Memory capacity constraints\newline\textbullet~Complex indexing optimization\\ 
\midrule

Long-term Memory and Continual Learning 
& \cite{wang2023augmenting,rae2020transformers,paissan2024structured,li2019learn}
& Utilizes memory-augmented neural architectures and lightweight continual learning for incremental knowledge updates
& \textbullet~Extended temporal reasoning\newline\textbullet~Avoids catastrophic forgetting\newline\textbullet~Continuous adaptation
& \textbullet~Memory management complexity\newline\textbullet~Performance degradation risks\\ 
\midrule

Causal and World-Model Prediction 
& \cite{hafner2019learning,zhang2020designing,cullen2023predicting}
& Implements causal reinforcement learning, latent dynamics modeling, and predictive simulation for proactive edge-based planning
& \textbullet~Enhanced decision safety\newline\textbullet~Reduced trial-and-error cost\newline\textbullet~Improved planning efficiency
& \textbullet~High model complexity\newline\textbullet~Sensitivity to inaccuracies\\ 

\bottomrule
\end{tabular}
\end{table*}

\section{Open Source Agentic AI Projects}

Agentic AI has proliferated in open-source environments, providing diverse, practical applications spanning various domains. Table~\ref{tab:Agentic_ai_github_summary} presents representative projects organized into three main categories: \textit{Agent Frameworks and Platforms}, \textit{Autonomous AI Agent Applications}, and \textit{Domain-specific AI Agents}.

\subsection{Agent Frameworks and Platforms}

Agent frameworks and platforms facilitate the deployment and management of autonomous, intelligent agents capable of reasoning, decision-making, and collaboration. They provide foundational tools enabling both developers and non-technical users to effectively harness Agentic AI for various practical scenarios, significantly lowering the barrier to entry for advanced agent-based applications~\cite{10757328}.

\textit{MetaGPT}: MetaGPT is a multi-agent collaborative framework utilizing natural language programming and task automation to facilitate efficient task execution among multiple autonomous agents. This framework is introduced and detailed in the paper by Hong et al.~\cite{hong2023metagpt}. Specifically, the authors proposed a sophisticated role-based architecture, empowering agents with distinct responsibilities such as autonomous code generation, peer code review, iterative refinement, and coordinated execution. This design significantly enhanced agents' collective problem-solving capabilities, effectively demonstrating the practical utility of Agentic AI in automating complex software engineering processes and minimizing human intervention.

\textit{Langflow}: Langflow is a low-code platform specifically designed for developing multimodal and retrieval-augmented generation (RAG)-based multi-agent systems. This framework is introduced and detailed in the paper by Jeong et al.~\cite{jeong2025beyond}. Specifically, the authors proposed visual and intuitive workflows, empowering agents with autonomous capabilities to process complex multimodal inputs (including text and images), dynamically orchestrate their interactions, and execute tasks without extensive coding. This design significantly enhanced autonomous agent collaboration, effectively demonstrating Langflow's practical utility in simplifying the adoption of sophisticated Agentic AI within enterprise environments.

\textit{SuperAGI}: SuperAGI is an intuitive and highly practical framework for rapidly deploying and managing autonomous AI agents. Although lacking direct academic publication~\cite{Superagi2023}, this platform distinctly emphasized real-world applicability, empowering agents with features such as rapid instantiation, comprehensive lifecycle management, and seamless scalability. This design significantly enhanced the agents' capability to autonomously execute, coordinate, and manage complex tasks, effectively demonstrating the practical realization of autonomous decision-making and efficient task orchestration across diverse operational scenarios.

\textit{AutoGen}: AutoGen is an innovative framework for developing complex applications through multi-agent conversations. This framework is introduced and detailed in the paper by Wu et al.~\cite{wu2023autogen}. Specifically, the authors proposed a highly adaptable conversational architecture, empowering agents with diverse tools, including human interactions, LLM-driven decision-making, and external service invocations. This design significantly enhanced agents' autonomous collaboration capabilities, effectively demonstrating AutoGen's strength in handling intricate workflows across various application domains, such as mathematics, coding, question-answering, and operational research.

\textit{AgentGPT}: AgentBench is a comprehensive benchmark designed to rigorously assess the capabilities of LLMs functioning as autonomous agents across multiple interactive environments. This benchmark is introduced and detailed in the paper by Liu et al.~\cite{liu2023agentbench}. Specifically, the authors proposed systematic evaluation methods, empowering agents with critical competencies such as autonomous reasoning, dynamic decision-making, iterative task-solving, and interactive tool utilization. Their results significantly highlighted performance disparities between commercial models (e.g., GPT-4) and open-source alternatives, effectively demonstrating the urgency for enhancing agent-oriented fine-tuning, training strategies, and robust open-source models explicitly tailored for autonomous agent applications.

\begin{table*}[!t]
\centering
\caption{Representative Open-Source Agentic AI Projects from GitHub}
\label{tab:Agentic_ai_github_summary}
{\fontsize{8.5pt}{7.5pt}\selectfont
\begin{tblr}{width=\linewidth,
  colspec={Q[2,c,m] Q[2,c,m] Q[4,l,m] Q[3.5,l,m] Q[2.5,c,m]},
  row{1}={c},
  hlines,
  vline{2-5}={-}{},
  vline{2}={2}{-}{},
}
\textbf{Task Domain} & \textbf{Project} & \textbf{Description} & \textbf{Key Feature} & \textbf{Repository Link} \\
\SetCell[c=5]{c}\textbf{Agent Frameworks and Platforms}\\
Agent Framework & MetaGPT & Modular multi-agent framework for collaborative task execution & Natural language programming, task automation & \url{https://github.com/geekan/MetaGPT} \\
Agent Orchestration & Langflow & Low-code pipeline builder for RAG and multi-agent systems & Visual workflow, intuitive orchestration & \url{https://github.com/logspace-ai/langflow} \\
Agent Management & SuperAGI & Framework for rapid deployment and management of autonomous agents & Agent lifecycle management & \url{https://github.com/TransformerOptimus/SuperAGI} \\
Agent Platform & AutoGen & Platform to build interactive, generative agent applications & Multi-agent communication, dynamic execution & \url{https://github.com/microsoft/autogen} \\
Agent Development & AgentGPT & Simplified GPT-based agent creation and management tool & Easy-to-use agent interface & \url{https://github.com/reworkd/AgentGPT} \\

\SetCell[c=5]{c}\textbf{Autonomous AI Agent Applications}\\
Software Engineering & OpenHands & Autonomous agent for production-level code generation & Autonomous planning, coding & \url{https://github.com/All-Hands-AI/OpenHands} \\
Collaborative AI & CrewAI & Role-based orchestration for cooperative AI agents & Task decomposition, collaboration & \url{https://github.com/joaomdmoura/crewAI} \\
Decision-making & AutoGPT & GPT-powered autonomous reasoning and self-improvement & Iterative decision-making & \url{https://github.com/Significant-Gravitas/AutoGPT} \\
Autonomous Coding & GPT-Engineer & Agent that autonomously generates complete software solutions & End-to-end automated coding & \url{https://github.com/AntonOsika/gpt-engineer} \\
Research Automation & ResearchGPT & AI agent for autonomous research and summarization & Autonomous information extraction & \url{https://github.com/mukulpatnaik/researchgpt} \\

\SetCell[c=5]{c}\textbf{Domain-specific AI Agents}\\
Cybersecurity & Real-time Threat Detection & Autonomous cybersecurity agent analyzing network traffic & Real-time network threat analysis & \url{https://github.com/OpenBMB/XAgent} \\
Autonomous Vehicles & Self-driving Delivery & Autonomous driving simulator integrating sensor fusion & Route planning, perception & \url{https://github.com/carla-simulator/carla} \\
Education & Virtual Tutoring & Adaptive personalized tutoring agent & Interactive, adaptive instruction & \url{https://github.com/huangwl18/VoxPoser} \\
Finance & FinGPT & Autonomous AI agent for financial data analysis and predictions & Financial forecasting, investment insights & \url{https://github.com/AI4Finance-Foundation/FinGPT} \\
Healthcare & BiMediX & Autonomous agent aiding medical diagnostics and healthcare research & Medical diagnostics, clinical decision support & \url{https://github.com/mbzuai-oryx/BiMediX} \\
\end{tblr}
}
\end{table*}

\subsection{Autonomous AI Agent Applications}

Autonomous AI agent applications enable agents to independently execute complex tasks through advanced reasoning, dynamic decision-making, and iterative task management. They significantly enhance productivity and effectiveness across specialized domains, showcasing the direct impact of Agentic AI technologies in practical scenarios.

\textit{OpenHands}: OpenHands is an autonomous agent designed for production-level code generation tasks. This framework is introduced and detailed in the paper by Selvaraj et al.~\cite{selvaraj2021openhands}. Specifically, the authors proposed an integrated system architecture empowering agents with capabilities such as autonomous planning, systematic coding, iterative refinement, and production-oriented software automation. This design significantly enhanced the agents' ability to autonomously generate high-quality code, effectively demonstrating the practical utility of Agentic AI in software engineering automation.

\textit{AutoGPT}: AutoGPT is an autonomous agent framework utilizing GPT models for iterative reasoning and self-improvement. This framework is introduced and detailed in the paper by Richards et al.~\cite{Significant_Gravitas_AutoGPT}. Specifically, the authors proposed a dynamic iterative reasoning loop, empowering agents with capabilities such as autonomous problem-solving, dynamic decision-making, continuous self-assessment, and refinement of strategies. This design significantly enhanced the agents' ability to autonomously handle diverse, complex tasks, effectively demonstrating the strength of Agentic AI in practical, adaptive scenarios.

\textit{CrewAI}: CrewAI is a role-based orchestration framework for cooperative AI agents. This framework is introduced in the open-source project by Moura et al.~\cite{crewAI2023}. Specifically, the authors proposed structured orchestration methods, empowering agents with clearly defined roles such as planners, researchers, executors, and coordinators. This design significantly enhanced collaborative problem-solving and task decomposition capabilities, effectively demonstrating practical utility in managing sophisticated workflows through autonomous agent cooperation.

\textit{GPT-Engineer}: GPT-Engineer is an autonomous coding agent designed to fully automate software solution generation. This framework is introduced in the open-source project by Osika et al.~\cite{Osika_gpt-engineer_2023}. Specifically, the authors proposed an autonomous pipeline that interprets user-defined requirements, autonomously designs software architectures, generates functional code, and iteratively refines the output. This design significantly enhanced end-to-end software development automation, effectively demonstrating Agentic AI’s capability in delivering rapid, reliable, and autonomous software engineering solutions.

\textit{ResearchGPT}: ResearchGPT is an autonomous agent designed to automate the research process comprehensively. This framework is introduced in the open-source project by Patnaik et al.~\cite{researchgpt2023}. Specifically, the authors proposed an autonomous research workflow empowering agents with capabilities such as systematic literature review, structured information extraction, summarization, and insightful synthesis. This design significantly enhances productivity and accuracy in complex research tasks, effectively demonstrating the practical utility of Agentic AI in automating rigorous academic and professional research activities.

\subsection{Domain-specific AI Agents}

Domain-specific AI agents are specialized autonomous systems explicitly designed to handle tasks unique to particular application areas. They leverage specialized domain knowledge and targeted capabilities to significantly enhance performance and practicality within specific operational contexts.

\textit{XAgent}: XAgent is an autonomous cybersecurity agent designed specifically for real-time network threat detection. This framework is introduced and detailed in the open-source project by OpenBMB~\cite{xagent2023}. Specifically, the authors proposed a robust autonomous monitoring system, empowering agents with capabilities for rapid threat identification, real-time security analysis, and dynamic cybersecurity responses. This design significantly enhanced network security effectiveness, effectively demonstrating the practical utility of Agentic AI in autonomous cybersecurity.

\textit{CARLA}: CARLA is an autonomous driving simulator designed explicitly for self-driving delivery tasks through comprehensive sensor fusion and realistic simulation scenarios. This framework is introduced and detailed in the paper by Dosovitskiy et al.~\cite{Dosovitskiy2017carla}. Specifically, the authors proposed a realistic urban environment simulation, empowering agents with sensorimotor control, adaptive scenario-driven evaluations, and robust navigation capabilities amidst dynamic obstacles, including other vehicles and pedestrians. This design significantly enhanced autonomous driving training, effectively demonstrating CARLA’s crucial role in practical self-driving applications.

\textit{VoxPoser}: VoxPoser is an adaptive personalized tutoring agent leveraging composable 3D value maps guided by language models for robotic manipulation tasks. This framework is introduced and detailed in the paper by Huang et al.~\cite{huang2023voxposer}. Specifically, the authors proposed integrating large language models to autonomously interpret and execute complex natural-language instructions, dynamically generating composable 3D affordance maps. This design significantly enhanced robotic manipulation capabilities, effectively demonstrating the practical application of Agentic AI in personalized and interactive educational environments.

\textit{FinGPT}: FinGPT is an autonomous AI agent explicitly developed for financial data analysis and predictive insights. This framework is introduced and detailed in the paper by Liu et al.~\cite{liu2023fingpt}. Specifically, the authors proposed democratizing internet-scale financial datasets through generative AI models, empowering agents with capabilities such as autonomous financial forecasting, investment decision support, and real-time financial analytics. This design significantly enhanced financial decision-making processes, effectively demonstrating FinGPT’s utility in intelligent financial services and investment management.

\textit{BiMediX}: BiMediX is an autonomous AI agent explicitly designed for bilingual medical diagnostics and clinical decision support. This framework is introduced and detailed in the paper by Pieri et al.~\cite{pieri2024bimedix}. Specifically, the authors proposed a bilingual MoE architecture, empowering agents with advanced medical diagnostic capabilities, clinical record analysis, and robust healthcare recommendations in multiple languages. This design significantly enhanced clinical decision accuracy and healthcare accessibility, effectively demonstrating BiMediX’s practical impact on intelligent and inclusive medical services.

\section{Case Studies of Agentic AI for Edge General Intelligence}

In this section, we present four representative applications of Agentic AI specifically tailored for edge general intelligence, i.e., low-altitude economy networking (LAENet), {intent networking}, {vehicular networks}, and {human-centric service provisioning}.

\subsection{Agentic AI for Low Altitude Economy Networking}
\begin{figure*}[t]
  \centering
  \includegraphics[width=0.95 \linewidth]{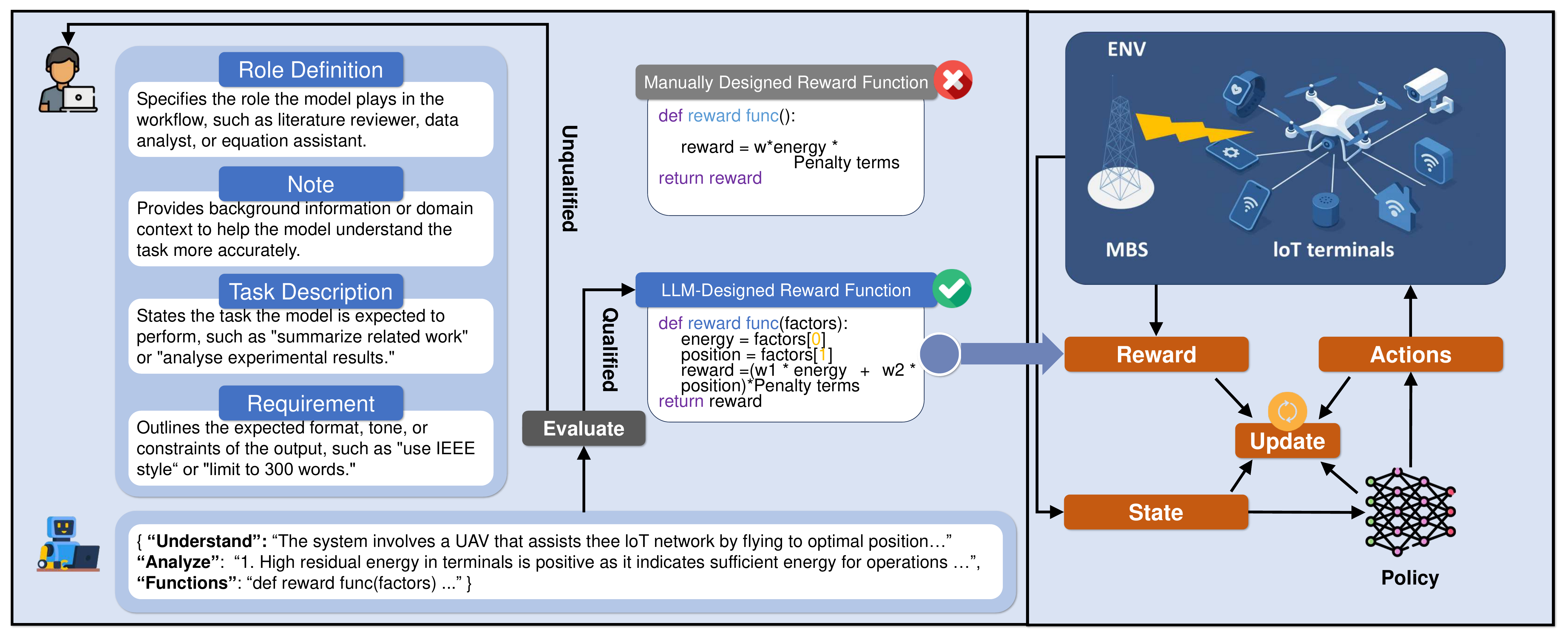}
  \caption{Architecture of a UAV-assisted IoT network with LLM-designed reward function for reinforcement learning in the LAENet framework. 
The system integrates compact LLMs for multimodal perception, structured reasoning, adaptive reward shaping, and decentralized coordination among UAV agents, enabling efficient and scalable data collection in edge environments~\cite{cai2025largelanguagemodelenhancedreinforcement}.}
  \label{fig:LLMASREWARDSYSTEMMODEL}
\end{figure*}

\subsubsection{Background and Motivation}  In the context of LAENet, supporting diverse aerial operations demands sophisticated real-time decision-making capabilities to cope with dynamic environments, stringent resource constraints, and heterogeneous network conditions \cite{2025cao,wei2025multi,zhao2025temporalspectrumcartographylowaltitude,gao2025agenticsatelliteaugmentedlowaltitudeeconomy,8368236}. Although RL has demonstrated significant promise for autonomous and adaptive aerial network control, classical RL methodologies frequently encounter severe limitations such as insufficient generalization to novel scenarios, suboptimal reward design, and unstable policy convergence, particularly in dynamic and uncertain aerial environments \cite{oubbati2022synchronizing,11049053,cheng2020comprehensive,jiang2023age}. For example, traditional RL methods struggle to adaptively adjust trajectories and energy-efficient operations for UAVs due to their fixed policy structures and simplistic reward designs, thereby hindering the practical applicability in complex real-world tasks~\cite{9701330,9769985,9558857}.

However, Agentic AI empowered by LLMs has emerged as a transformative paradigm, integrating advanced cognitive functions such as contextual understanding, dynamic generalization, and structured reasoning, thereby substantially enhancing autonomous decision-making~\cite{2025cao,zhang2024generative,liu2022blockchain}. Unlike traditional RL methods, Agentic AI leverages pretrained LLMs to extract multimodal features, enabling contextually adaptive reward shaping and action selection. In particular, COT prompting allows LLMs to effectively capture contextual nuances and reason through complex scenarios, significantly improving generalization across heterogeneous aerial tasks~\cite{wei2022chain}. LLMs support task decomposition and plan revision through both forward and backward reasoning, enabling agents to adaptively solve complex problems with interpretable steps~\cite{ren2024thinking}. Furthermore, LLM-based reward shaping has demonstrated superior task alignment and stability compared to manually designed reward functions~\cite{kwon2023reward}. Moreover, agents can coordinate via shared LLMs by exchanging abstract intents and jointly planning actions, achieving decentralized collaboration without explicit protocols~\cite{qiu2024collaborative}. Integrating these sophisticated cognitive capabilities into the RL loop thus overcomes traditional RL's inherent limitations, providing a robust and scalable solution for adaptive and efficient LAENet deployments.


\subsubsection{System Description} As depicted in Fig.~\ref{fig:LLMASREWARDSYSTEMMODEL}, we consider a UAV-assisted IoT communication network within the framework of LAENet, comprising a single UAV, a macro base station (MBS), and multiple distributed IoT terminals~\cite{cai2025largelanguagemodelenhancedreinforcement,qu2020dynamic}. In this scenario, the UAV maintains a fixed altitude and constant cruising speed, dynamically adjusting its hovering positions near the IoT terminals to optimize data collection and energy delivery. The terminals leverage harvested energy wirelessly provided by the UAV to transmit sensor data, which the UAV subsequently aggregates and forwards to the MBS for further processing. However, maintaining optimal hovering positions to ensure data throughput and reliability presents a critical trade-off: continuous UAV repositioning significantly escalates propulsion and communication energy consumption, complicating the optimization of system energy efficiency.

Under these operational considerations, we formulate an aerial data collection and energy efficiency multi-objective optimization problem aiming to minimize total system energy consumption including terminal transmission energy, UAV propulsion, and communication energy, while satisfying stringent constraints on transmission power limits, data throughput requirements, decoding reliability, and data freshness. This optimization problem inherently features high-dimensional, non-convex, and NP-hard characteristics due to dynamic environmental factors and real-time constraints~\cite{Ullah2025DRL}. Classical optimization techniques typically decompose such problems into separable convex subproblems solved iteratively; however, the effectiveness of these approaches heavily relies on decomposition strategies and faces severe computational overhead in dynamic IoT environments~\cite{li2025unauthorized}. Agentic AI offers adaptive decision-making capabilities and sophisticated contextual reasoning without explicit decomposition~\cite{sapkota2025aiagentsvsagentic,oubbati2022synchronizing}. By embedding LLM-generated adaptive reward signals directly into RL frameworks, Agentic AI effectively navigates complex state-action spaces, achieving robust, scalable, and near-optimal solutions for UAV localization and energy allocation, thus significantly outperforming traditional optimization methods in dynamic LAENet scenarios.

\subsubsection{Workflow of Agentic AI framework for LAENet} Generally, the Agentic AI framework substantially enhances the adaptive reasoning and policy optimization capabilities by effectively integrating the contextual comprehension and structured reasoning strengths of LLMs with the sequential decision-making capacity of RL. Specifically, the workflow of Agentic AI for LAENet is structured into four key stages~\cite{cai2025largelanguagemodelenhancedreinforcement}, addressing complex decision-making scenarios involving multimodal inputs and dynamic environmental conditions. Here, we elaborate the workflow through an illustrative UAV-assisted IoT data collection scenario in LAENet to demonstrate the efficacy of the integrated approach.

\begin{itemize}
    
    \item \textit{Step 1: State Perception and Abstraction:} The UAV–environment interaction is formulated as a Markov decision process (MDP), where the state includes information such as UAV location, residual energy, and channel conditions. Agentic AI leverages pretrained LLMs to perceive and abstract these heterogeneous inputs. To support edge deployment, lightweight LLM variants (e.g., LoRA-adapted or quantized models) are used to encode multimodal sensory signals and user instructions efficiently. This yields compact yet expressive state representations that facilitate robust decision-making.
    \item \textit{Step 2: Action Selection and Policy Execution:} Based on the perceived state, the LLM guides action generation by dynamically reasoning over possible trajectories. Specifically, chain-of-thought prompting enables the decomposition of high-level objectives into subgoals, improving transparency and adaptability. LLMs further perform causal reasoning to anticipate the outcomes of sequential actions, enabling more informed policy execution. Compact actor-critic networks or distilled policy modules are employed to meet real-time execution constraints under energy and bandwidth limitations.

    \item \textit{Step 3: Reward Evaluation and Feedback Processing:} During execution, the agent collects both explicit feedback (e.g., sensed delay) and implicit feedback (e.g., human-in-the-loop comments). Agentic AI uses LLMs to interpret such signals and adaptively construct reward functions aligned with mission goals. Compared to manually designed functions, the adaptive reward shaping mechanism better accommodates environmental variability and user preferences. It also enables context-aware trade-offs between data freshness and resource consumption.
    \item \textit{Step 4:Policy Update and Knowledge Integration:} The collected trajectories and reward feedback are used to iteratively refine the RL policy. The LLMs summarize episodic knowledge and integrates it into a continually evolving policy. In multi-agent scenarios, agents exchange intent summaries and local knowledge via shared LLM-based communication protocols, enabling decentralized coordination for collaborative coverage, scheduling, and resource sharing. To maintain efficiency and scalability, memory-efficient architectures such as RAG-based retrieval modules are adopted for long-term knowledge reuse.
\end{itemize}

By embedding advanced Agentic AI capabilities, such as multimodal comprehension, dynamic context adaptation, and structured reasoning, into every stage of the RL decision-making loop, the LLM-enhanced RL framework markedly improves agent intelligence, adaptability, and interpretability. Consequently, this approach provides a robust, scalable, and human-aligned solution for secure, autonomous, and adaptive UAV-assisted IoT data collection and operation in complex and dynamic LAENet environments.
\begin{figure}[t]
  \centering
  \includegraphics[width=1\linewidth]{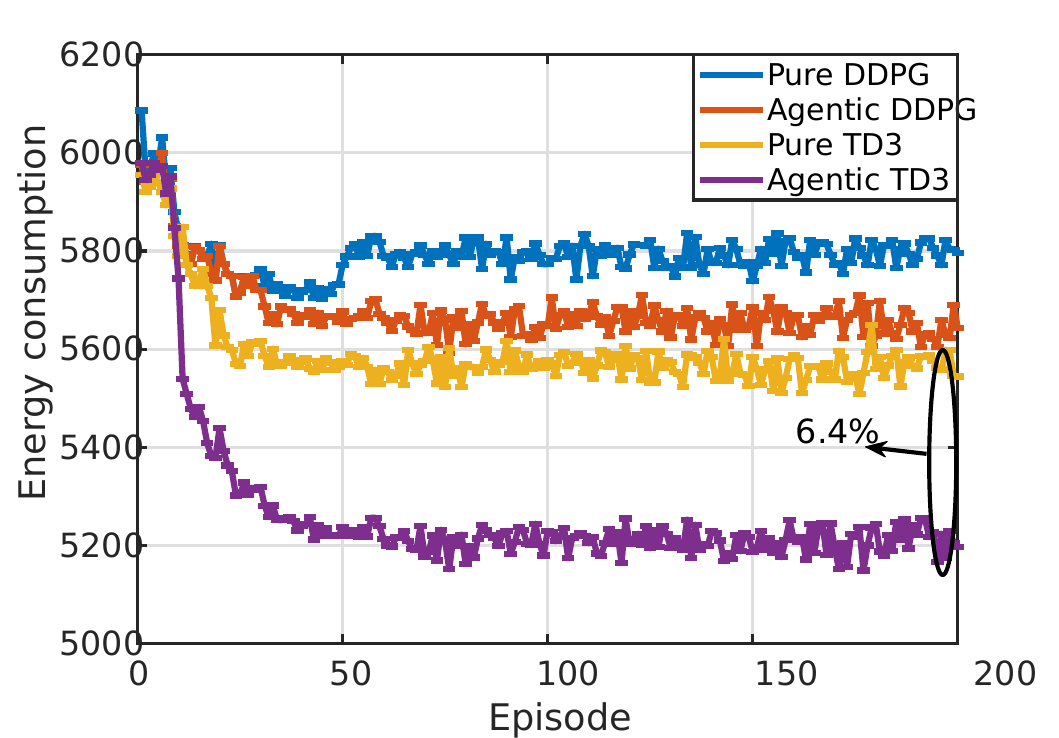}
  \caption{Energy consumption across episodes for various algorithms using Pure DRL versus Agentic DRL~\cite{cai2025largelanguagemodelenhancedreinforcement}.}
  \label{fig:LLM_reward}
\end{figure}

\subsubsection{Numerical Results} Fig.~\ref{fig:LLM_reward} presents the convergence performance comparison of the proposed Agentic AI-enhanced reward design approach with conventional manually designed rewards for DDPG and TD3 algorithms. Notably, algorithms equipped with LLM-generated rewards demonstrate consistently superior performance, achieving substantial reductions in total energy consumption. Specifically, the Agenitc TD3 attains up to a 6.4\% reduction in final energy consumption compared to its manually designed counterpart. This performance enhancement can be primarily attributed to the richer reward structure generated by the LLM, which incorporates comprehensive UAV positional information alongside energy-related factors. Consequently, this enables the UAV to dynamically optimize its trajectory, effectively reducing flight distances and communication overhead.

Additionally, the effectiveness of the Agentic AI-enhanced reward design indicates promising generalization potential for more intricate and diverse optimization tasks, such as multi-objective or cross-domain resource allocation scenarios in LAENet~\cite{wei2025multi,li2025unauthorized}. Conversely, traditional DRL methods constrained by manually crafted rewards exhibit limited performance and flexibility~\cite{oubbati2022synchronizing,10529221}, failing to sufficiently adapt to the real-time variability and complexity inherent to LAENet environments.

\subsubsection{Lessons Learned} 
Agentic AI-driven RL effectively incorporates high-level cognitive reasoning and contextual comprehension provided by LLMs~\cite{2025cao,zhang2024generative}, enabling robust and adaptive decision-making within complex, dynamic environments. The integration of pretrained LLM-generated adaptive reward mechanisms fundamentally transforms traditional reward design approaches, generating nuanced, context-aware reward signals and action selections~\cite{kwon2023reward}. This significantly mitigates classical DRL limitations, including suboptimal local convergence and rigid exploration strategies. Consequently, Agentic AI not only resolves critical challenges identified in conventional RL methods, such as inadequate generalization and unstable policy convergence, but also demonstrates substantial potential for addressing more sophisticated and multidimensional optimization tasks, particularly in complex multi-objective or cross-domain LAENet scenarios~\cite{li2025unauthorized,9267779}.

\begin{figure*}[t]
  \centering
  \includegraphics[width=0.95 \linewidth]{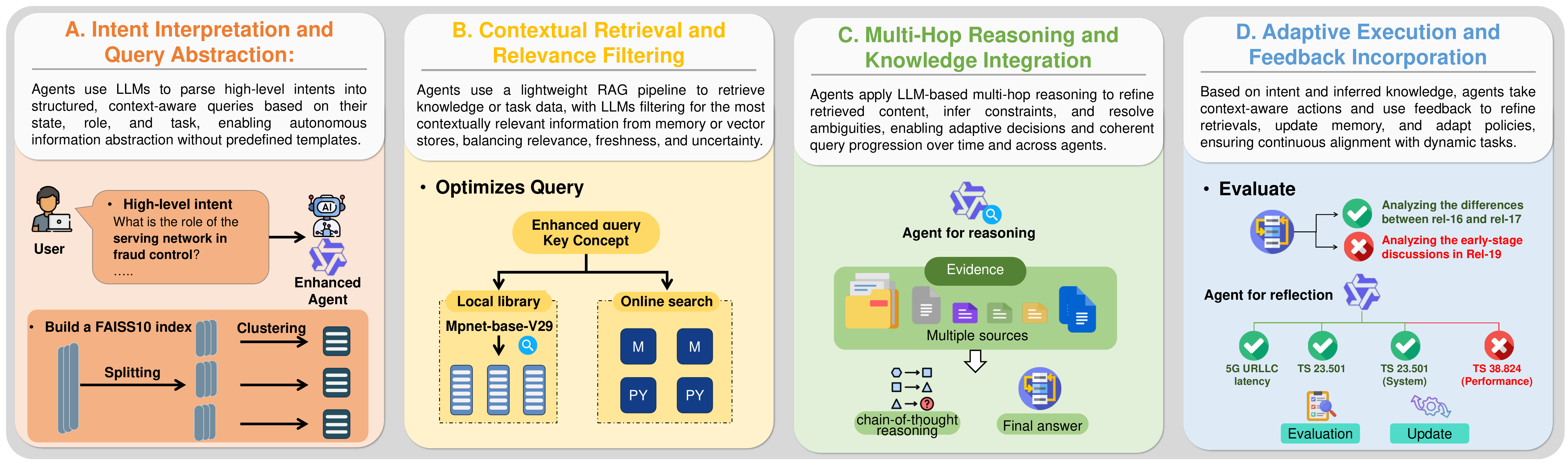}
  \caption{Illustration of the Agentic contextual retrieval enhanced intelligent base station for troubleshooting and decision-making~\cite{zhang2025agenticaigenerativeinformation}.}
  \label{fig:Intent_network_sys}
\end{figure*}

\subsection{Agentic AI for Intent Networking}
\subsubsection{Background and Motivation}
In next-generation intelligent networking systems, context-aware knowledge retrieval has become essential for enabling timely, relevant, and adaptive decision-making across dynamic and resource-constrained environments~\cite{liu2025lametaintentawareagenticnetwork}. Traditional retrieval-augmented networking architectures often rely on centralized indexing or fixed rule-based matching mechanisms, which limit scalability and responsiveness under rapidly changing network states, such as in vehicular networks, aerial relays, or multi-agent swarms. These methods typically fail to support online semantic reasoning, multi-hop task tracking, or fine-grained spatial-temporal alignment, thereby impairing the system’s ability to deliver high-quality contextual information across diverse and evolving scenarios.

Agentic offers a transformative paradigm by introducing dynamic, in-situ retrieval capabilities that align with the agent's internal decision-making context~\cite{2025cao,zhang2024generative}. Unlike static retrieval frameworks, Agentic AI enables edge agents to autonomously interpret natural language queries, reason over latent task histories, and proactively retrieve or generate semantically relevant content based on mission objectives. For instance, recent advances in RAG allow agents to access distributed knowledge bases and refine results through iterative interactions~\cite{han2023comprehensive}. Furthermore, chain-of-thought prompting enables structured, multi-step reasoning over retrieved content, allowing agents to infer, filter, and apply contextual cues in real time~\cite{wei2022chain}. By embedding such capabilities into communication-aware systems, Agentic AI fundamentally augments network intelligence, enabling semantic query routing, proactive data fusion, and context-driven protocol adaptation. This integration not only enhances agent collaboration and responsiveness but also paves the way for scalable, memory-efficient, and knowledge-grounded networking infrastructures suited for real-world edge deployments.

\subsubsection{System Description}
As illustrated in Fig.~\ref{fig:Intent_network_sys}, we consider an intent network system empowered by Agentic AI, where distributed edge agents are tasked with interpreting high-level user intents and autonomously translating them into actionable, network-wide behaviors. In such environments, agents must operate under conditions of limited observability, dynamic topologies, and heterogeneous device capabilities~\cite{zhang2025agenticaigenerativeinformation}. Traditional intent translation pipelines, often rule-based or statically programmed, lack the flexibility to adapt to evolving network states or to reason over ambiguous or under-specified intents, thereby limiting responsiveness and scalability.

To address these limitations, we propose an Agentic AI framework in which each network agent is equipped with a compact LLM to support real-time semantic understanding, contextual reasoning, and adaptive intent interpretation. Agents collaborate via multi-hop communication and utilize contextual prompts to retrieve relevant policy templates, network state information, and domain knowledge from distributed knowledge bases using RAG mechanisms~\cite{han2023comprehensive}. This allows agents to resolve intents dynamically based on current network conditions and task history, rather than relying on predefined intent-to-policy mappings. The system aims to minimize intent translation latency and maximize execution accuracy while preserving scalability and autonomy. This is achieved by optimizing retrieval granularity, knowledge routing strategies, and response composition through LLM-driven reasoning. Compared to static intent network architectures, the Agentic AI-based system demonstrates superior adaptability, enabling network agents to proactively infer user goals, disambiguate conflicting intents, and generate context-aware action plans without centralized orchestration. This architecture provides a scalable and human-aligned solution for intent realization in next-generation edge-native intelligent networks~\cite{zhang2024generative,2025cao}.

\subsubsection{Workflow of Agentic Contextual Retrieval Framework}

Generally, the Agentic Contextual Retrieval (ACR) framework leverages the cognitive and semantic capabilities of LLMs to empower networked agents with proactive, goal-aligned information retrieval and intent grounding. Unlike static or rule-based retrieval systems, ACR enables autonomous agents to dynamically interpret, decompose, and fulfill high-level intents in situ by integrating RAG, distributed memory access, and structured reasoning. Specifically, the ACR workflow is structured into four stages that collectively support scalable and adaptive intent-driven operations across network agents~\cite{zhang2025agenticaigenerativeinformation}.

\begin{itemize}
    \item \textit{Step 1: Intent Interpretation and Query Abstraction:} Upon receiving a high-level intent (e.g., ``ensure full-area coverage within 10 minutes''), the agent formulates structured semantic queries based on its local state, role, and contextual task awareness. This involves LLM-powered parsing of natural language into symbolic or task-grounded representations, enabling agents to autonomously abstract context-specific information needs without pre-defined templates.
      \item \textit{Step 2: Contextual Retrieval and Relevance Filtering:} The agent issues a retrieval prompt, either locally or across peers, via a lightweight RAG pipeline to access knowledge entries, cached task traces, or environmental facts. Through embedded attention and filtering mechanisms, the LLM identifies the most contextually relevant entries from distributed memory buffers or vector stores, balancing relevance with freshness and uncertainty.

    \item \textit{Step 3: Multi-Hop Reasoning and Knowledge Integration:} Retrieved content is iteratively processed using LLM-based multi-hop reasoning to infer higher-order relationships, resolve ambiguity, or refine the query outcome. For example, coverage plans may be adapted by inferring constraints from recent UAV paths, network congestion, or peer statuses. The agent synthesizes this information into actionable decisions or next-hop queries, ensuring continuity of reasoning across time and agents.

    \item \textit{Step 4: Adaptive Execution and Feedback Incorporation:} Based on the interpreted intent and inferred knowledge, the agent executes appropriate actions (e.g., adjusting trajectory or reassigning coverage roles), and monitors environmental and inter-agent feedback. This feedback is then used to refine future retrievals, update memory indices, and guide policy evolution, thereby enabling continual alignment between evolving intent expressions and dynamic task realities.

\end{itemize}

By embedding Agentic AI capabilities, such as semantic grounding, distributed memory reasoning, and intent-aware retrieval, into every phase of the information acquisition loop, the ACR framework transforms passive data access into a proactive, interpretive, and self-evolving process. It enables agents to collaboratively fulfill intents in uncertain, bandwidth-limited, and partially observable environments, laying a robust foundation for scalable and context-adaptive networking intelligence.

\subsubsection{Numerical Results}

\begin{figure}[t]
  \centering
  \includegraphics[width=0.49\textwidth]{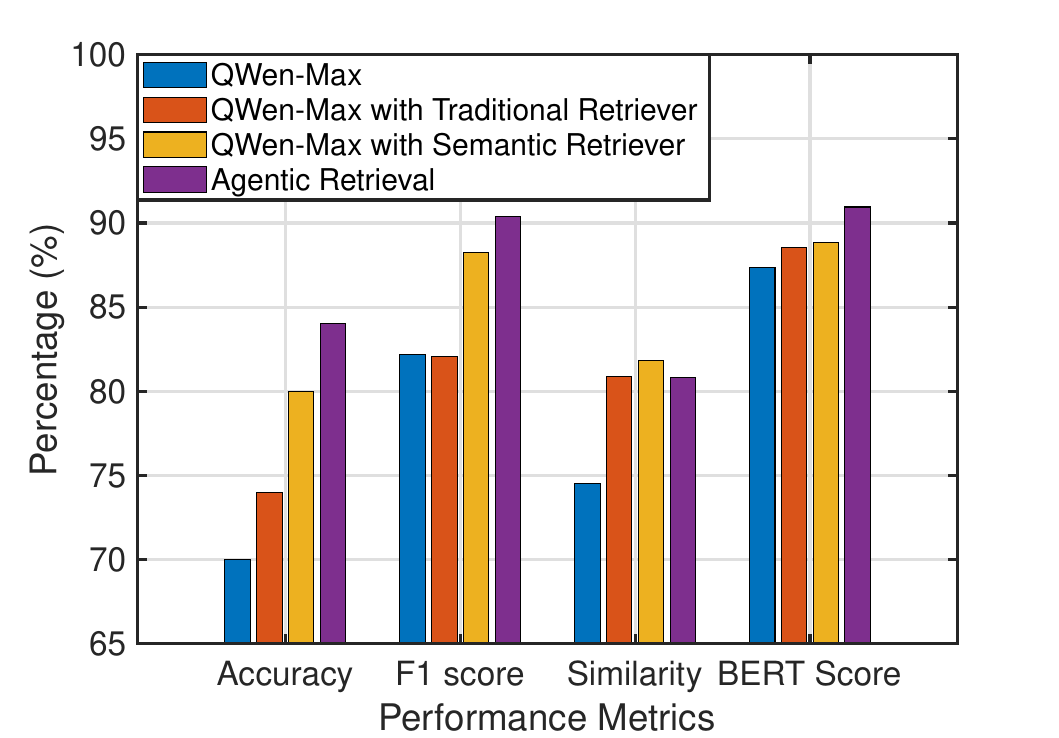}
  \caption{Comparison of Agentic Retrieval performance with baseline methods, including QWen-Max without a retriever, traditional retrieval, and semantic retrieval~\cite{zhang2025agenticaigenerativeinformation}.}
  \label{fig:contextual_retrieval_performance}
\end{figure}

 Fig.~\ref{fig:contextual_retrieval_performance} presents the performance comparison between the proposed ACR framework and conventional query matching baselines under varying intent complexities. The ACR approach, empowered by LLM-driven semantic interpretation and reasoning, achieves higher intent fulfillment accuracy and faster response convergence across all scenarios. Specifically, under complex multi-agent intents involving conditional constraints and partial observability, ACR improves task success rate by up to 14.8\% compared to traditional keyword-based or rule-based retrieval methods. This performance gain is attributed to the ability of Agentic AI agents to interpret natural language intents, reason over distributed memory, and iteratively refine retrieval prompts based on contextual cues. Furthermore, by leveraging RAG mechanisms and lightweight in-situ LLMs, agents adaptively prioritize relevant knowledge entries and suppress redundant query broadcasts, yielding up to 23.4\% communication reduction compared to uniform broadcast schemes. This efficiency stems from Agentic AI’s capacity to align retrieval actions with both task-specific objectives and environmental context, rather than treating retrieval as an isolated or static subroutine. Additionally, the results demonstrate validate that embedding Agentic AI capabilities into retrieval workflows enables scalable, goal-aligned, and context-aware information access~\cite{ReAct}. This lays a foundation for robust and semantically grounded intent resolution in future networked intelligence systems~\cite{Shen2024EI,He2024TMC}.

\subsubsection{Lessons Learned}
The Agentic Contextual Retrieval framework illustrates the significant advantages of embedding Agentic AI capabilities into intent-based networking architectures~\cite{zhang2024generative,2025cao}, effectively addresses longstanding challenges in traditional intent networks, including rigid intent-to-policy mappings, static retrieval logic, and limited adaptability to evolving task contexts. The integration of RAG mechanisms and in-situ LLM inference enables agents to align retrieval strategies with user goals and environmental dynamics, improving both responsiveness and interpretability~\cite{gupta2024comprehensivesurveyretrievalaugmentedgeneration}. These capabilities not only enhance intent fulfillment accuracy and communication efficiency but also lay the groundwork for scalable, distributed intelligence in real-world, partially observable environments~\cite{yang2020edgeintelligenceautonomousdriving}. Ultimately, Agentic AI offers a transformative approach to enabling self-adaptive, goal-driven collaboration across network agents, establishing a practical and extensible foundation for the next generation of intent-aware edge-native networking systems~\cite{9119487}.

\subsection{Agentic AI for Vehicular Edge Computing}
\subsubsection{Background and Motivation}
Mobile Edge Computing (MEC) has emerged as a key enabler of low-latency, high-throughput services in dynamic vehicular environments. However, traditional MEC frameworks often rely on static scheduling policies or centralized decision logic, which struggle to scale under high mobility, variable wireless links, and user heterogeneity~\cite{you2017energy,mao2017survey,7932863}. As vehicular networks evolve toward ultra-dense deployments and semantically rich applications (e.g., autonomous driving, cooperative perception), MEC must go beyond computation offloading and align system resources with user intent and context~\cite{8584062,8245844}.

By embedding autonomous, intent-aware agents within edge nodes (i.e., vehicles), Agentic AI enables the system to gain the capacity that is parse natural-language objectives, perceive semantic environment signals, and dynamically coordinate offloading or scheduling decisions. This study presents a representative Agentic AI framework for edge computing in vehicular systems, where vehicles act as embodied agents integrating semantic inference (via LLAVA~\url{https://github.com/haotian-liu/LLaVA}) and adaptive decision-making (via GAE-PPO~\cite{zhang2025embodiedaienhancedvehicularnetworks}), aligned with perceived user intent through the Weber-Fechner-inspired QoE model~\cite{11049053}.

\subsubsection{System Description}

\begin{figure*}[t]
\centering
\includegraphics[width=0.95\textwidth]{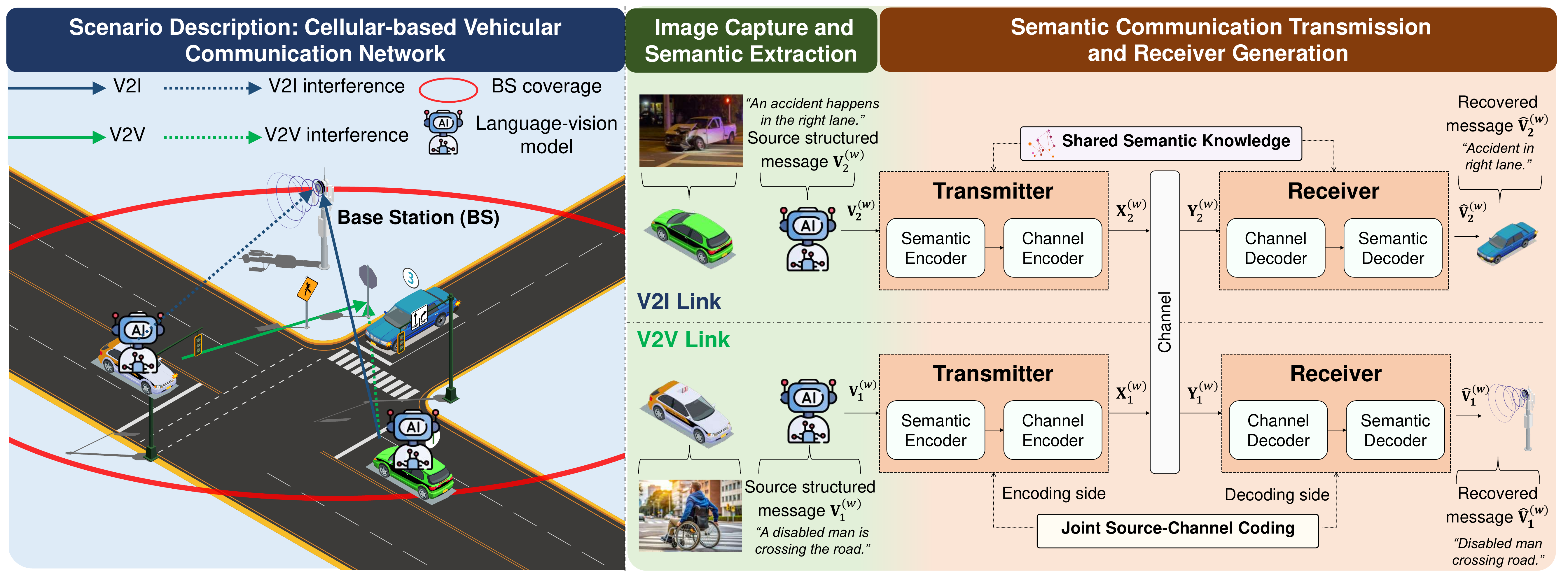}
\caption{{System model illustrates a cellular-based vehicular communication network, where embodied AI vehicles utilize semantic communication to encode and decode structured messages for efficient and reliable data exchange \cite{Shunpu_SemCom}.}}
\label{fig:enter-label--}
\end{figure*}

As illustrated in Fig.~\ref{fig:enter-label--}, we consider a cellular-based vehicular edge computing system in which $I$ vehicles operate as embodied agents equipped with onboard AI processors and cameras. The network supports V2I and V2V communications over $W$ subbands and includes a base station responsible for coarse-grained spectrum coordination~\cite{zhang2025embodiedaienhancedvehicularnetworks}. Each vehicle captures environmental images and uses the LLAVA model to extract semantic information (e.g., object descriptions, parking availability). The information is encoded and transmitted to infrastructure or peers using a semantic communication stack. The offloading and scheduling decisions are modeled as a joint optimization problem aiming to maximize a Weber-Fechner-based QoE metric subject to SINR, symbol length, power, and semantic similarity constraints. These constraints embed both network-level resource feasibility and user-level perceptual utility, forming a context-rich decision space where Agentic AI agents operate.

\subsubsection{Workflow of Agentic AI for MEC}
The agentic AI framework for mobile edge task scheduling and transmission control contains the following steps.

\begin{itemize}
    \item \textit{Step 1: Intent Interpretation and Semantic Abstraction:}  
    Upon observing raw visual inputs from the surrounding environment, each vehicle utilizes LLAVA to extract structured semantic representations that encapsulate objects, spatial layouts, and driving context. These semantic outputs are aligned with implicit user intents and serve as the basis for intent-grounded policy generation.

    \item \textit{Step 2: Policy Retrieval and Decision Generation:}  
    The semantic intent vector is mapped to a latent task profile, which is either matched against previously successful policies stored in distributed memory or processed through an online GAE-PPO decision module. This yields a set of adaptive action parameters, including transmission power level, selected communication channel, and semantic symbol length, all optimized under current environmental and network constraints.

    \item \textit{Step 3: Constrained Execution and QoE-Aware Evaluation:}  
    Based on the selected policy, semantic messages are encoded and transmitted through V2V or V2I links. The receiving node reconstructs the message and evaluates its semantic fidelity via cosine similarity between BERT-based embeddings of the original and decoded text. This quality signal forms the basis for assessing the user-perceived effectiveness of the transmission.

    \item \textit{Step 4: Feedback Integration and Policy Refinement:}  
    The agent computes a reward signal that integrates semantic accuracy and transmission cost using a Weber-Fechner-inspired QoE function~\cite{10654734}. This reward is used to update the policy network via GAE-PPO, enabling continual improvement of intent-grounded behavior over time. Additionally, performance traces are stored for future retrieval, closing the learning loop.
\end{itemize}
By embedding Agentic AI capabilities, such as semantic abstraction, context-driven policy generation, and reward-aligned adaptation, into each stage of the edge decision-making loop, the proposed framework transforms traditional scheduling into a cognitively enriched, intent-responsive process. It empowers mobile agents to reason over multimodal observations, align actions with human-perceived utility, and continuously refine behavior in real-time. This design lays the foundation for scalable, human-aligned, and semantically adaptive mobile edge intelligence.

\subsubsection{Numerical Results}
\begin{figure}[!t]
\centering
\includegraphics[width=0.49\textwidth]{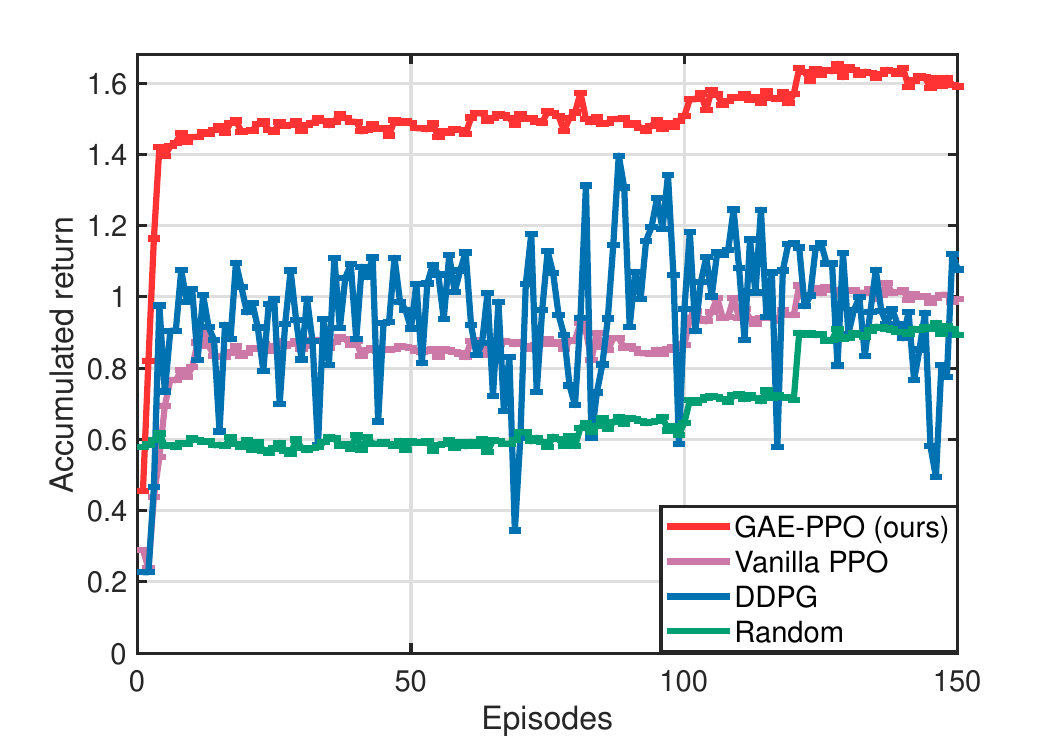}
\caption{Convergence behavior with different methods~\cite{Shunpu_SemCom}.}
\label{fig:enter-label-3}
\end{figure}

Fig.~\ref{fig:enter-label-3} shows the convergence behavior of the Agentic AI-enabled method in comparison with several baseline algorithms, including pure PPO, DDPG, and a random policy. It achieves consistently higher returns per episode and exhibits significantly faster convergence and specifically outperforms pure PPO by a margin of approximately 61\% in accumulated return, highlighting its superior sample efficiency and stability. Collectively, these results validate that the Agentic AI framework by embedding GAE into the actor-critic learning loop, achieves more reliable and sample-efficient policy optimization, rendering it well-suited for adaptive decision-making in mobile edge vehicular networks.

\subsubsection{Lessons Learned}
The Agentic AI framework for MEC demonstrates the practical benefits of embedding LLM-driven semantic reasoning and reinforcement-based policy optimization into dynamic vehicular environments. It effectively overcomes critical limitations of conventional MEC systems, including static resource scheduling, task-agnostic transmission, and lack of real-time adaptability to user-level goals. By integrating LLAVA-based semantic abstraction with GAE-PPO-enhanced policy evolution, the framework enables autonomous agents to align communication and computation strategies with perceived task intent and environmental conditions~\cite{11049053,9575181}. These capabilities not only improve decision stability and semantic transmission efficiency, but also promote self-adaptive and perceptually grounded coordination among mobile edge nodes~\cite{khelloufi2025agi}. Ultimately, this case study demonstrates that Agentic AI provides a scalable and context-aware paradigm for intent-aligned task scheduling and resource optimization in future edge-native intelligent systems~\cite{10505907,10090477,8926369}.

\begin{figure*}[t]
\centering
\includegraphics[width=0.95\textwidth]{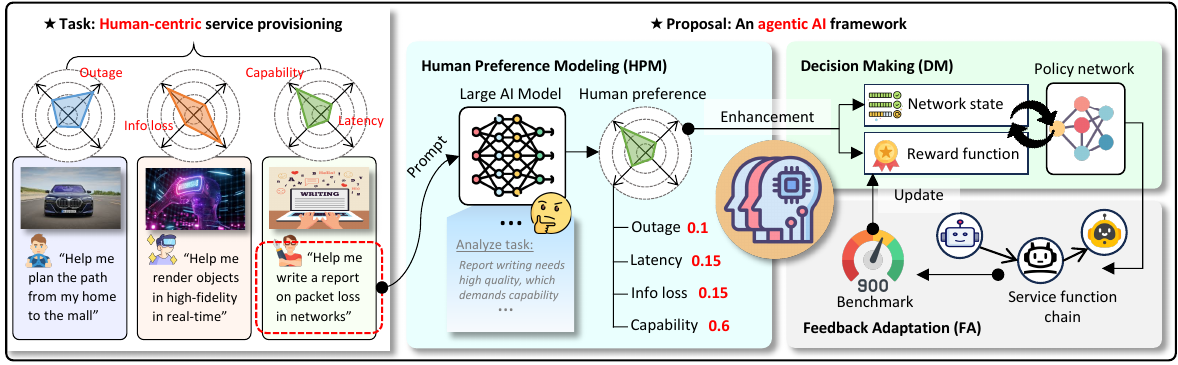}
\caption{{The illustration of an agentic AI framework for human-centric service provisioning in Edge General Intelligence  \cite{liu2025lametaintentawareagenticnetwork}.}}
\label{fig:enter-label}
\end{figure*}

\subsection{{Agentic AI for Human-centric Service Provisioning}}

\subsubsection{Background and Motivation}
EGI aims to serve human users with personalized and context-aware services across diverse application domains. However, traditional edge intelligence systems predominantly focus on generic optimization objectives, such as minimizing latency or maximizing throughput, often neglecting subjective preferences and contextual requirements that define the human-centric service experience~\cite{10144631,torres2020immersive,zhuang2019sdn}. 
The fundamental challenge in human-centric service provisioning lies in the difficulty of translating subjective human preferences into actionable optimization strategies for Service Function Chain (SFC) composition \cite{liu2025lametaintentawareagenticnetwork,qin2025generative,zhou2025user}. Existing approaches rely on predefined QoE metrics that fail to capture the nuanced, context-dependent nature of human perception and satisfaction. For instance, different users may prioritize different aspects of service quality: some users may emphasize capability and generation quality for content creation tasks, while others may prioritize low latency and reliability for real-time applications~\cite{qoe1, qoe2,li2021cost}.

Agentic AI can transform edge systems by enabling them to autonomously understand natural-language human preferences, dynamically optimize SFC compositions, and adapt proactively through continuous learning~\cite{liu2025lametaintentawareagenticnetwork,krishnan2025ai}. Unlike conventional edge intelligence that operates with static optimization targets, Agentic AI leverages LLMs to interpret diverse expressions of user satisfaction, translates these into structured preference vectors, and employs a DRL-based planning module to optimize the SFC composition dynamically~\cite{gao2025agenticsatelliteaugmentedlowaltitudeeconomy,yang2025frontiers}. This cognitive approach transforms edge general intelligence from resource-centric optimization to truly human-centric service provisioning, maximizing subjective QoE.

\subsubsection{System Description}
As shown in Fig.~\ref{fig:enter-label}, we present an agentic AI framework to perform human-centric service provisioning. The  edge general intelligence environment comprises multiple distributed edge servers, each providing specific services (e.g., content generation, data analysis, and multimedia processing). Moreover, we employ a Centralized Large AI Model (C-LAM) at the cloud infrastructure and multiple lightweight Edge Large AI Models (E-LAMs) distributed across edge servers. The C-LAM serves as a central coordinator that maintains comprehensive user preference databases, while edge servers host multiple E-LAMs with varying model sizes and preference understanding capabilities to serve users with different computational budgets and latency requirements. Users submit service requests that require multi-step SFC composition, where each step involves selecting an appropriate service provider from various candidates.

The proposed Agentic AI framework integrates three core technological components: Human Preference Modeling (HPM), Decision Making (DM), and Feedback Adaptation (FA). 
The HPM module captures and quantifies subjective human preferences through advanced knowledge distillation techniques, where the C-LAM continuously monitors user interactions and transfers preference understanding capabilities to lightweight E-LAMs via weighted pairwise distillation processes \cite{liu2025lametaintentawareagenticnetwork}. 
Each E-LAM learns to interpret natural language expressions of user satisfaction and contextual cues, translating them into structured preference vectors that capture relative priorities for different service quality dimensions. 
The DM module integrates preference-guided reasoning with DRL to optimize SFC composition and resource allocation, formulating SFC compositions as an optimization problem. 
The FA module enables continuous system improvement by collecting multi-modal feedback from users and incorporating this information to refine both preference understanding and decision making, ensuring that the Agentic AI framework adapts to evolving user requirements and contextual conditions.

\subsubsection{Workflow of Agentic AI for Human-centric Service Provisioning}
The proposed framework operates through a three-stage workflow that seamlessly integrates human preference understanding with adaptive SFC optimization.

\begin{itemize}
\item \textit{Step 1: Human Preference Interpretation:} Users submit service requests through natural language prompts that contain explicit task descriptions, implicit quality expectations, and contextual information such as urgency levels, resource constraints, and task-usage scenarios. Then, an E-LAM processes this multimodal input along with the current environmental context to generate a personalized preference vector $\mathbf{s} = [\omega_C, \omega_B, \omega_L, \omega_P]$ that quantitatively represents the user's relative priorities for service capability, information fidelity, response latency, and system reliability. This preference interpretation leverages chain-of-thought reasoning to decompose complex user requirements into quantifiable dimensions.

\item \textit{Step 2: Preference-Guided SFC Composition:} The DM module is based on a DRL architecture that formulates service provisioning as a Markov Decision Process. Particularly, the current network state, incorporating network conditions, resource availability, and agent capabilities, is augmented with preference-weighted features to form the enhanced state representation. The preference vector $\mathbf{s}$ simultaneously modulates the reward function design, ensuring that the learning process is guided toward maximizing user-perceived quality rather than generic system metrics. This preference-aware DRL enables the policy network to generate optimal SFC compositions that cater to specific users.

\item \textit{Step 3: Feedback Integration and System Adaptation:} The selected SFC is executed on the distributed edge infrastructure, while the system continuously monitors both objective performance metrics and subjective satisfaction indicators derived from user behaviors. An in-context learning mechanism is integrated into FA, where the E-LAM maintains a structured context memory containing historical records of user preferences, generated SFCs, and resulting satisfaction outcomes. This contextual memory enables FA to detect preference patterns, adapt to evolving user requirements, and implement automatic calibration mechanisms when preference misalignment is detected. The continuous feedback loop ensures that both the preference understanding capabilities and the DRL policies are refined through accumulated user interactions, achieving symbiotic enhancement between cognitive reasoning and adaptive optimization.
\end{itemize}

This integrated workflow enables the transformation of traditional resource-centric edge optimization into a truly human-centric, adaptive service provisioning system that continuously evolves to better serve individual user needs while maintaining system efficiency and scalability.
\begin{figure}[!t]
\centering
\includegraphics[width=0.49\textwidth]{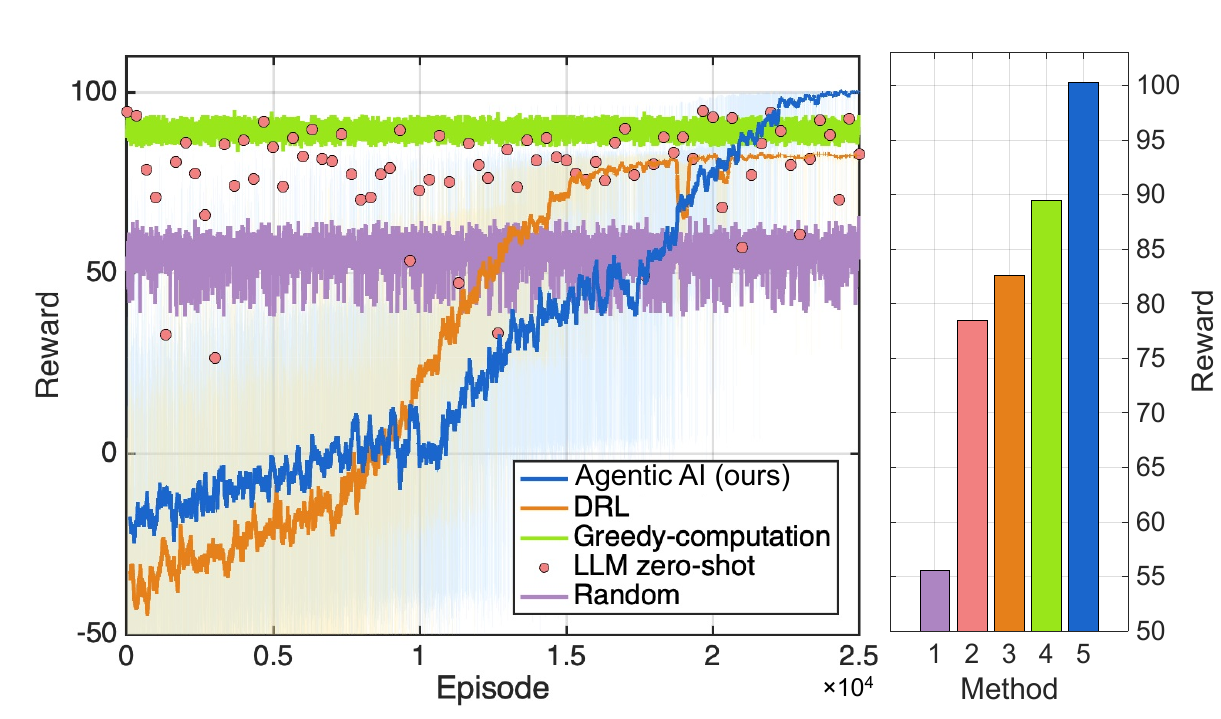}
\caption{The performance and learning curves of different methods in human-centric service provisioning~\cite{liu2025lametaintentawareagenticnetwork}.}
\label{fig:enter-label3}
\end{figure}

\subsubsection{Numerical Results}
Fig.~\ref{fig:enter-label3} shows the performance comparison of the proposed Agentic AI framework against conventional optimization baselines for human-centric service provisioning.
We evaluate the framework on a representative application scenario involving personalized content generation services, where users submit requests for generating technical reports, creative content, and data analysis outputs with diverse quality expectations and contextual requirements. 

The experimental results demonstrate that the Agentic AI framework achieves consistently superior performance across all evaluation metrics. Specifically, our preference-aware approach attains up to 27.3\% improvement in human-centric QoE compared to traditional DRL methods that assume uniform user preferences. The superior performance can be attributed to the synergistic integration of E-LAM preference interpretation with preference-guided DRL optimization. Unlike conventional approaches that optimize for generic system metrics, our framework dynamically adapts optimization objectives according to individual user preference vectors, enabling more precise alignment between system behavior and human expectations.

\subsubsection{Lessons Learned}
This case demonstrates a paradigm shift from traditional resource-centric optimization toward truly cognitive and human-centric edge general intelligence. The revolutionary advantage of Agentic AI lies in the natural language understanding and contextual reasoning capabilities introduced by LLMs, along with the decision-making and feedback modules constructed around LLMs, enabling edge systems to autonomously interpret subjective human preferences, dynamically adapt optimization objectives according to individual user contexts, and continuously evolve through accumulated user interactions \cite{liu2025lametaintentawareagenticnetwork,ferrag2025llmreasoningautonomousai,huang2024digital,11007275}. This work establishes a foundation for future research directions, including multi-stakeholder preference reconciliation, privacy-preserving personalization, and long-term preference evolution modeling.

\section{Future Research Directions}

Emerging research directions focus on the synergistic advancement of Agentic AI and edge general intelligence to effectively address the complex demands and inherent constraints of next-generation edge networks~\cite{luo2025edgegeneralintelligencemultiplelarge,jiang2025large,10929033}. Key future avenues emphasize the integration of cognitive autonomy, resource efficiency, robust decision-making, and seamless adaptability in real-world operational contexts. These directions include:

\begin{itemize}

    \item \textbf{Adaptive and Efficient Collective Intelligence:}
    Investigating scalable frameworks for decentralized agent collaboration to enhance cognitive autonomy within resource-constrained edge general intelligence deployments. Research should develop efficient decentralized consensus methods, adaptive task allocation strategies, and robust emergent communication mechanisms, enabling AI agents to autonomously collaborate and adapt via agentification process in heterogeneous edge environments~\cite{8951115,10908978}. Moreover, future systems must dynamically adjust collaboration granularity and communication frequency based on network congestion, agent density, and environmental volatility.

    \item \textbf{Privacy-Preserving Federated Agent Systems:}
    Developing federated learning methodologies tailored explicitly for Agentic AI and edge general intelligence scenarios, emphasizing scalable, privacy-preserving model training and deployment. Research should advance secure aggregation protocols, adaptive federated architectures, and decentralized knowledge sharing techniques, facilitating collective agent intelligence while maintaining stringent data privacy requirements~\cite{rodríguezbarroso2025challengestrustworthyfederatedlearning,10960683,luo2025trustworthy}. Federated optimization should further accommodate heterogeneous agent capabilities and unreliable communication links common in edge environments.

    \item \textbf{Robustness and Safety in Autonomous Reasoning:}
    Designing robust frameworks to ensure reliable and transparent decision-making capabilities for Agentic AI systems within dynamic edge general intelligence contexts. Future studies should explore real-time hallucination detection methods, autonomous validation of reasoning outputs, causal interpretability techniques, and fail-safe operational mechanisms, enabling trustworthy performance in critical edge applications such as autonomous vehicles and smart manufacturing~\cite{11029984,karim2025largelanguagemodelsapplications,luo2024bc4llm}. Furthermore, formal verification and self-diagnostic modules should be integrated to monitor reasoning integrity in mission-critical deployments.

    \item \textbf{Cross-Domain Adaptation and Migration:}
    Developing effective methods for Agentic AI systems to seamlessly generalize knowledge and adapt across diverse operational scenarios typical of edge general intelligence environments. Research should focus on robust cross-domain transfer techniques, efficient knowledge migration strategies, and adaptive learning frameworks, allowing agents to autonomously adjust to varying contexts without significant retraining overhead~\cite{10993375,10197260,10614179}. Memory-based transfer mechanisms, self-supervised domain alignment, and continual learning under resource-aware constraints are promising directions.

    \item \textbf{Compression-Aware agentification Reasoning:}
    Investigating compression-aware architectures specifically designed to integrate explicit reasoning capabilities into resource-constrained edge systems. Future research should focus on the co-design of model compression techniques, such as low-rank adaptation, structured pruning, quantization, and knowledge distillation, with advanced agentification reasoning mechanisms, ensuring cognitive expressiveness, real-time responsiveness, and energy efficiency at the edge~\cite{wang2025empowering,chen2025transforminghybridcloudemerging,zheng2025review}. In particular, hierarchical modular designs and dynamic sparsification could enable reasoning-aware compression with minimal performance degradation.

\end{itemize}

\section{Conclusion}

This paper has provided a comprehensive survey of Agentic AI and agentification process tailored explicitly for edge general intelligence. It has systematically introduced foundational concepts and clearly distinguished Agentic AI from traditional edge intelligence paradigms. Key enabling technologies, including model compression, energy-aware computing, robust connectivity, and knowledge representation and reasoning methods, have been reviewed. Representative Agentic AI applications such as LAENet, intent-driven networking, vehicular networks, and human-centric service provisioning have been illustrated through detailed case studies and experimental analyses. Additionally, this survey has discussed critical deployment challenges, examined emerging open-source frameworks, and identified promising directions for future research.


\bibliographystyle{IEEEtran}
\bibliography{IEEEabrv.bib,REF}

\end{document}